\documentclass[useAMS,usenatbib]{mn2e}
\usepackage{BibDef}

\usepackage[fleqn]{amsmath}
\usepackage{graphicx}
\usepackage{multirow}
\usepackage{natbib}
\usepackage{color}

\title[Definition of environment in a hierarchical Universe]
  {The definition of environment and its relation to the quenching of galaxies at z=1-2
  in a hierarchical Universe.}
\author[M.Fossati et al.]
  {M.~Fossati,$^{1,2}$\thanks{E-mail: mfossati@mpe.mpg.de} D.J.~Wilman,$^{1,2}$ F.~Fontanot,$^3$ G.~De~Lucia,$^3$ P.~Monaco,$^{4,3}$
  \newauthor 
  M.~Hirschmann,$^3$ J.T.~Mendel,$^2$ A.~Beifiori,$^{1,2}$ E.~Contini$^3$ \\
  \\
  $^1$Universit{\"a}ts-Sternwarte M{\"u}nchen, Schenierstrasse 1, D-81679 M{\"u}nchen, Germany.\\
  $^2$Max-Planck-Institut f{\"u}r Extraterrestriche Physik, Giessenbachstrasse, D-85748 Garching, Germany.\\
  $^3$INAF - Osservatorio Astronomico di Trieste, via G.B. Tiepolo 11, I-34131 Trieste, Italy \\
  $^4$Dipartimento di Astronomia dell'Universit{\'a} di Trieste, via G.B. Tiepolo 11, I- 34131 Trieste, Italy }
  
\date{Accepted 2014 October 27.  Received 2014 October 20; in original form 2014 March 28}

\pagerange{\pageref{firstpage}--\pageref{lastpage}} \pubyear{2000}

\def\LaTeX{L\kern-.36em\raise.3ex\hbox{a}\kern-.15em
    T\kern-.1667em\lower.7ex\hbox{E}\kern-.125emX}

\begin{document}

\label{firstpage}

\maketitle

\begin{abstract}
  A well calibrated method to describe the environment of galaxies at
  all redshifts is essential for the study of structure formation.
  Such a calibration should include well understood correlations with halo
  mass, and the possibility to identify galaxies which dominate their
  potential well (centrals), and their satellites.  
  Focusing on z$\sim 1$ and $2$ we propose a method of
  environmental calibration which can be applied to the next
  generation of low to medium resolution spectroscopic surveys. 
  Using an up-to-date semi-analytic model of galaxy formation, we
  measure the local density of galaxies in fixed apertures on different scales.  
  There is a clear correlation of density with
  halo mass for satellite galaxies, while a significant population of 
  low mass centrals is found at high densities in the neighbourhood 
  of massive haloes.
  In this case the density simply traces the
  mass of the most massive halo within the aperture.
  To identify central and satellite galaxies, we apply an
  observationally motivated stellar mass rank method which is both
  highly pure and complete, especially in the more massive haloes where
  such a division is most meaningful.  Finally we examine a test case
  for the recovery of environmental trends: the passive fraction of
  galaxies and its dependence on stellar and halo mass for
  centrals and satellites.  With careful calibration, 
  observationally defined quantities do a good job of recovering known
  trends in the model. This result stands even with reduced 
  redshift accuracy, provided the sample is deep enough to preserve 
  a wide dynamic range of density.

\end{abstract}


\section{Introduction}
In the widely accepted Lambda-Cold Dark Matter ($\rm \Lambda CDM$)
scenario \citep{white+78,perlmutter+99} of
structure formation, primordial density fluctuations collapse into
virialized haloes. Baryons follow the gravitational field of the dark
matter giving birth to galaxies which then interact and merge. As a
result, galaxies can live in a great variety of different environments
possibly impacting their evolution and fate.  Indeed, the
fraction of passive galaxies, and the fraction of morphologically early-type galaxies,
show strong, positive correlations with the local (projected) density
of neighbouring galaxies (\citealt{dressler+80, balogh+97, balogh+04,
poggianti+99, kauffmann+04}; see also \citealt{blanton+09} for a review). 
A variety of stripping effects can and do act on galaxies in dense environments 
(see \citealt{boselli+06} for a review), for example ram pressure
stripping \citep{gunn+72, abadi+99}, strangulation \citep{larson+80}, 
and tidal stripping \citep{keel+85, dekel+03, diemand+07, villalobos+13}. 
However, the correlation of galaxy properties with environment can also reflect
differences in the merger and growth history of (particularly
massive) galaxies, driving correlations with halo, stellar and bulge
mass but also (indirectly) with density \citep[e.g.][]{woo+13, wetzel+13,
wilman+13, hirschmann+14}.

At $z \sim 1-2.5$, galaxies are in a stage of maximum growth
via star formation \citep{madau+96, elbaz+07, rodighiero+11} and mergers. 
To first order, this happens because gas and galaxies track the accretion 
of dark matter \citep[see e.g.][]{lilly+13, saintonge+13} and thus is a consequence 
of rapid early structure growth. Thus most satellites at this epoch are
experiencing a dense surrounding medium for the first time, and gas
stripping can have a dramatic impact on the suppression of active star
formation.

However, even at low redshift, where a wealth of multi-wavelength data
is available, it is non trivial to disentangle the relative
contribution of different physical processes as a function of stellar
mass, environment and the hierarchy of a galaxy in its own halo (e.g.
being a \textit{central} or a \textit{satellite}). It is therefore
clear that witnessing the rise and decline of the cosmic star formation
activity and its dependencies on environment in the early stages of the
life of the Universe is even more challenging.  Significant effort has
been invested in {\it testing} measurements of density in the face of
difficult selection and redshift errors at $z\sim1$ -- including survey
edge effects, magnitude selection, incompleteness and limited redshift
accuracy \citep{cooper+05, kovac+10}.  These efforts, and the
interpretation of resulting correlations of environment with galaxy
properties, are hampered by the inhomogeneity of methods used 
for different surveys and by the lack of calibration
to theoretically important parameters such as halo mass.
 
In the past couple of decades, efforts to model the evolution of
galaxies by pasting simple recipes describing baryonic physics onto the
hierarchical merger trees of dark matter (DM) haloes in a $\Lambda$CDM
Universe (semi-analytical models of galaxy formation, SAMs) have gained
momentum and had some success in reproducing the properties of the
galaxy population, especially at $z\sim0$ \citep[see e.g.][]{white+91,
  kauffmann+93, cole+94, cole+00, somerville+99, bower+06, croton+06,
  delucia+07, monaco+07, guo+11, guo+13}.  Although the predictions of
this class of model are still in tension with some observational
properties -- such as the evolution of low-mass galaxies
\citep{fontanot+09, weinmann+12, henriques+13}, the properties of
satellite galaxies \citep{weinmann+09, boylan+12, hirschmann+14}; the
baryon fraction on galaxy clusters \citep{mccarthy+07} -- SAMs can be
useful to calibrate and test methods to define environment at
  different redshift.

Our goal is twofold: First, we aim at defining a self-consistent and purely 
  observational parameter space within which the detailed dependence of 
  galaxy properties on their surrounding structure can be evaluated without 
  prejudice. We will thus use mock galaxy catalogues to construct a projected
  density field evaluated in redshift space, in order to test the impact of
  different definitions of density and at different redshift accuracy.
  We also compute a stellar mass rank within an aperture, which
  provides a purely observational parameter relating to the local
  gravitational dominance of a particular galaxy. These simple
  parameters contrast with complementary methods such as the
  construction of a group catalogue which forces each galaxy into
  a single halo, using an algorithm with idealized
  parameters derived from models.

Second, once such trends are established, it is equally important
  to examine how this contrasts with physical predictions and
  understand those trends in the context of theoretically important
  parameters such as halo mass and whether a galaxy is a central or
  satellite of its halo. We calibrate our observational parameters by 
  examining how they correlate with those accessible in a semi-analytic 
  model.

This two-step process is Bayesian in nature (galaxies have a well
  defined observational parameter set, while the theoretical parameter
  calibrations are probabilistic) which is well suited to statistical
  studies, as well as to the application of selection functions
  and measurement errors when simulating a real survey. 
Our definition for what we call ``environment'' can be equally applied at
high or low redshift, although in this paper we focus on high redshift
where new opportunities are beginning to open up with low
  resolution spectroscopic surveys conducted in the NIR
\citep[e.g.][]{brammer+12}. We also concentrate this paper
  exclusively on the calibration of environment using models and
  testing our recovery of known trends in the model using our methods.
  These methods will be applied to observational data in future
  papers.

There are many ways to describe the density field, e.g. number of
  neighbours within an adaptive or fixed cylindrical aperture, adaptive
  smoothing \citep{park+07}, Voronoi tessellation \citep{scoville+13} and shape
  statistics \citep{dave+97}. 
  
We focus on a set of simple density measurements using neighbouring galaxies.
  This is straightforward to correct for incompleteness and calibrate for 
  survey selection and for redshift errors.
There are typically two flavours of this method. The first,
based on the Nth nearest neighbour \citep{dressler+80, baldry+06,
  cooper+06, poggianti+08, brough+11} correlates only weakly with halo
mass \citep{haas+12, muldrew+12}.  The second, more sensitive to
high-mass over densities, and possibly easier to interpret, is based on
the number of galaxies within a fixed aperture
\citep{hogg+03, kauffmann+04, croton+05, gavazzi+10, wilman+10}.
Recently, \citet{shattow+13} have demonstrated that the fixed apertures
method is more robust across cosmic time, is less sensitive to the
viewing angle, and closer to the real over density measured in 3D space
than the Nth nearest neighbour. For those reasons, we use this method 
to quantify the environment.

The paper is structured as follows. In Section \ref{Models} we
introduce the semi-analytical models used and
we discuss the sample selection and the corrections needed to obtain a
sample comparable to observations.  In Section \ref{methodENV} we
present our method describing how to compute the local galaxy density.
In Section \ref{Sect_Halomassenv} we present a detailed analysis of
correlations between density defined on different scales and halo
masses, with a focus on the different behaviour of centrals and
satellites. In Section \ref{RankCenSat} we present an observationally
motivated method to identify centrals and satellites and we carefully
assess the performances of this method using the models.  In Section
\ref{passivefracvsenv} we test to what extent the tools we have
developed are effective in recovering environmental trends on physical
properties for model galaxies. We focus on a single property: the
fraction of passive galaxies because of its strong environmental
dependence in the models. Finally in Section \ref{conclusions} we
discuss and summarise the main results of this work.

Throughout the paper we assume a $\rm \Lambda CDM$ cosmology with the
following values derived from WMAP7 \citep{komatsu+11} observations:
$\Omega_m =0.272,~\Omega_b = 0.045,~\Omega_\Lambda =
0.728,~n=0.961,~\sigma_8 = 0.807 $ and $h = 0.704$.

\section{Models} \label{Models}
In this study we make use of the latest release of the Munich model as introduced by \citet[][hereafter G11]{guo+11} and later updated to 
WMAP7 cosmology by \citet[][hereafter G13]{guo+13}. This model takes advantage of a new run of the Millennium N-body 
simulation (Thomas et al. in prep.) which includes $N=2160^3$ particles within a comoving box of size $500h^{-1}$ Mpc on a side 
and cosmological initial conditions consistent with WMAP7 observational constraints. These are in reasonable agreement with the most recent 
result from both the WMAP9 \citep{hinshaw+12} and the Planck \citep{planck+13} missions. 
The particle mass resolution is $9.31 \times 10^8 h^{-1} M\odot$, and simulation data are stored at 62 output times. 
The most significant difference between the assumed cosmological parameters and those
used in the original Millennium run (which assumes WMAP1 cosmology) is a lower value of $\sigma_8$. This implies a lower amplitude 
for primordial fluctuations which is partially compensated by a higher value of $\Omega_m$. G13 performed a detailed comparison of 
several statistical properties of the galaxy population in WMAP1 and WMAP7 cosmologies. The impact of a lower $\sigma_8$ on the
galaxy population was also studied in detail in \citet[][using WMAP3 cosmology]{wang+08}, and we refer the reader to these papers for the 
details. G13 model is therefore optimally suited for our purposes due to the combination of the correct cosmology and the large volume.

In order to evaluate the local density around each galaxy, accurate positions of each sub-halo (the main unit hosting a galaxy) 
are crucial. In G13 main haloes are detected using a standard friends-of-friends (FOF) algorithm. Then each group is decomposed 
into sub-haloes running the algorithm \textsc{\small SUBFIND} \citep{springel+01}, which determines the self-bound structures 
within an FOF group. Each sub-halo hosts a galaxy. As time goes by, the model follows dark matter haloes 
after they are accreted onto larger structures. When two haloes merge, the galaxy hosted in the more massive halo is considered
the central, and the other becomes a satellite. After infall the haloes experience tidal truncation and stripping \citep{delucia+04, gao+04}
and their mass is reduced until they fall below the resolution limit of the simulation (20 bound particles, i.e. $2.64\times 10^{10} M_\odot$).
When this happens the sub-halo is no longer present in the simulation catalog but the galaxy still lives in the main halo 
(those objects are called orphan galaxies). Its lifetime is set by the dynamical friction formula \citep[see][]{delucia+10} 
and its position is assigned to the position of the most-bound particle in the sub-halo (this particle being defined at the last time 
the sub-halo was detected). 
Once this time has elapsed the galaxy is assumed to merge with the central galaxy of the main halo. Although crude, this assumption 
reproduces well some observational results like the radial density profile of galaxy clusters (which are dominated by orphan galaxies in
the central regions, \citealt{gao+04}), and the clustering amplitude on small scales \citep{wang+06}. 

The G13 model is based on earlier versions of the Munich model, e.g.
\citet{croton+06} and \citet{delucia+07}. It includes prescriptions for
gas cooling, star formation, size evolution, stellar and AGN feedback,
and metal enrichment.  G13 assumes a new and more realistic
implementation (compared to the simple model by \citealt{mo+98}) for
the sizes of gas discs.  Both supernovae (SN) and active galactic
nuclei (AGN) feedback are implemented.  The first implies that massive
stars explode as supernovae and the energy released converts a fraction
of the cold gas into the hot phase or even expels it from the halo. AGN
feedback is implemented following the model by \citet{croton+06}.  It
is assumed to be caused by `radio mode' outflows from a central black
hole that reduces and can completely suppress the cooling of hot gas
onto the galaxy. For more details on these prescriptions we refer the
reader to G11, G13 and \citet{delucia+07}.
\begin{figure*}
\begin{center}
\begin{tabular}{c c c}
\hspace{2cm} \includegraphics[scale = 0.30]{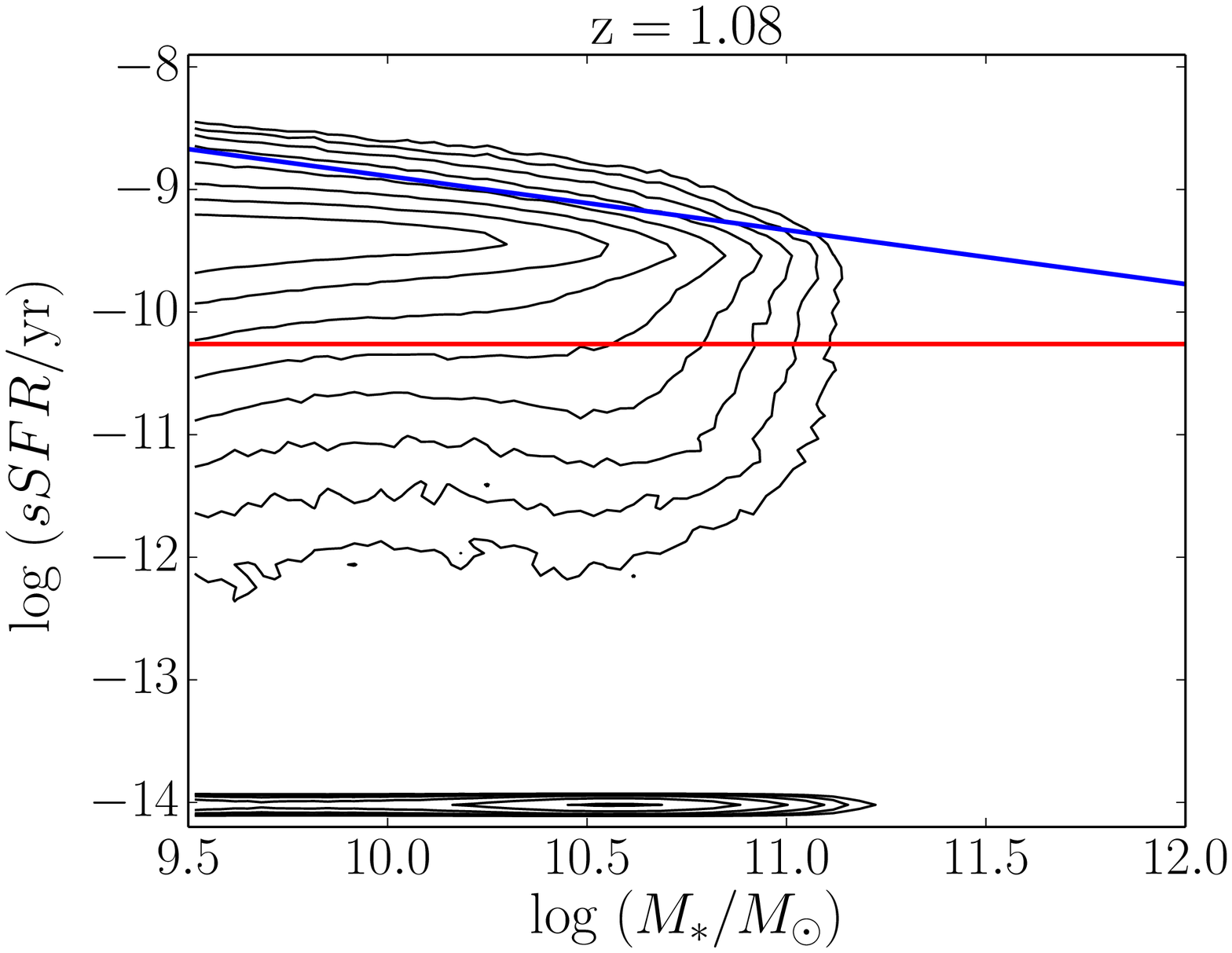} & \hspace{0.5cm} \includegraphics[scale = 0.30]{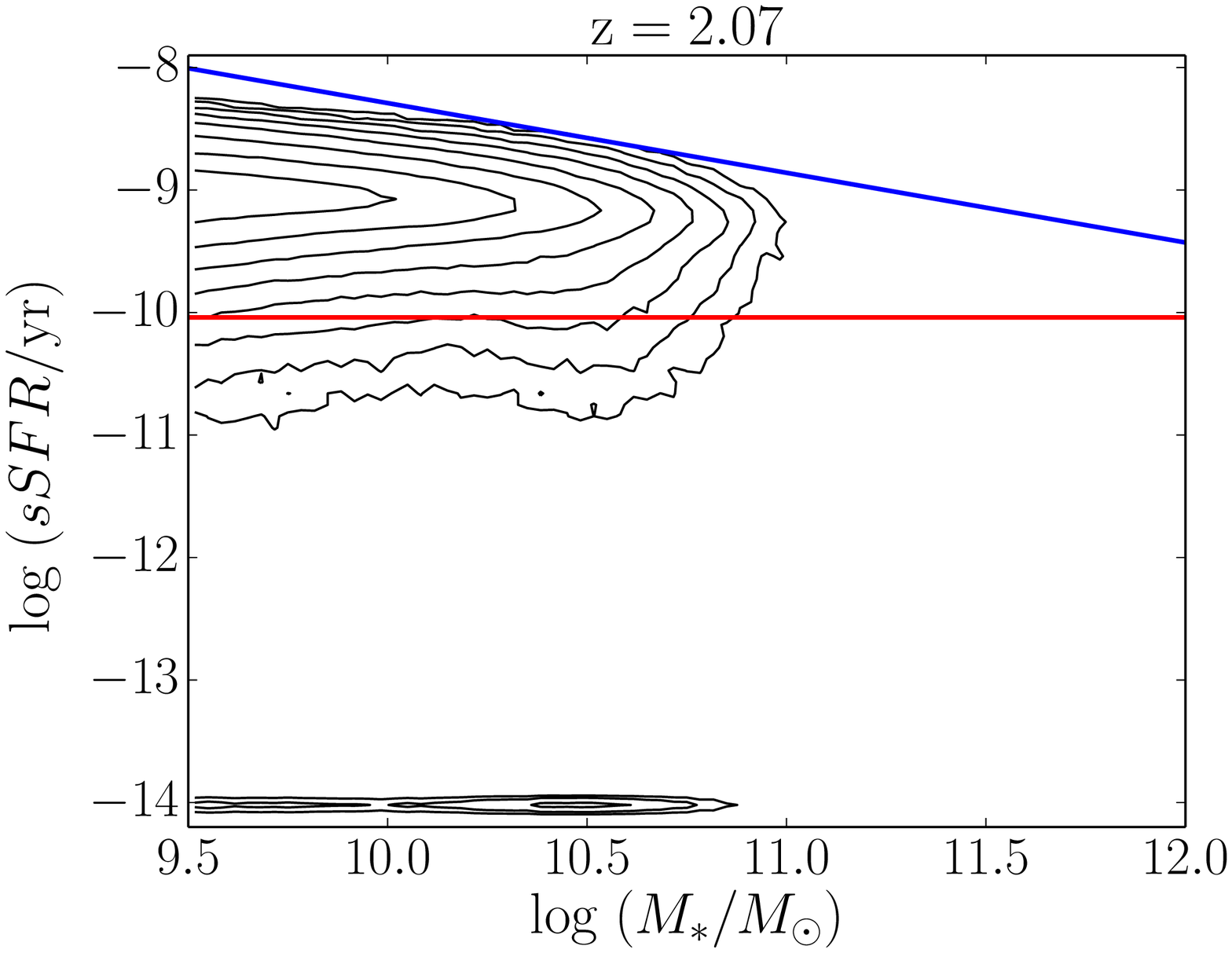} \\
\end{tabular}
\end{center}
\caption{Specific star formation rate ($sSFR$) as a function of $M_*$
  at redshifts of $1.08$ (left panel) and $2.07$(right panel).
  The blue line is the main sequence fit from observational data (NMBS,
  \citealt{whitaker+12}).  The red line marks the limit we set to
  define the passive galaxies in the models.  The contour levels are
  log-spaced with the outermost contour at 25 objects and the innermost at
  $10^4$ per bin. We set all galaxies with $\log sSFR < -14$ equal to that value.}
\label{sSFRMs}
\end{figure*}

It is worth mentioning that current SAMs suffer from an overproduction of low-mass galaxies at high-redshift. 
Recently, \citet{henriques+13} presented a new model in which the reincorporation timescales of galactic wind ejecta are a function
of halo mass. As a result they obtain a better fit of observed stellar mass functions out to $z \sim 3$. However, because this new 
prescriptions are not implemented in G13 we perform a statistical correction to the number densities as presented in Sect. \ref{ModelSEL}.
 
In most SAMs the satellite population has been quenched too quickly due to instantaneous stripping of the hot gas 
\citep[e.g.][]{weinmann+06, weinmann+10}. G11 proposed a more gentle action of strangulation and ram pressure stripping which 
are active only when a galaxy falls into the virial radius of a more massive halo. Although this improves the treatment of 
environmental effects, the fraction of passive galaxies is still significantly overestimated compared to observational 
results at $z=0$ as shown by \citet{hirschmann+14}. While this discrepancy is reduced in G13, the passive fraction is still 
too large.

\subsection{The model galaxy sample} \label{ModelSEL}
From the 62 outputs of the simulation, we make use of those at $z= 1.08$ and $z=2.07$.  
We select galaxies above a fixed stellar mass limit of $10^{9.5} M_\odot$ in both the redshift snapshots.
This limit is high enough to protect us against resolution bias in the models and on the other hand is as deep as 
the current spectroscopic observations at these redshift can realistically reach. 
Setting the same mass limit at different redshifts allows us to witness the number density increase 
as the Universe evolves and we get $\log(n/\rm{Mpc^3})= -2.08, -2.28$ at $z= 1.08, 2.07$ respectively.
Those values are higher than the number densities recently obtained by \citet{muzzin+13} integrating the 
stellar mass functions (SMFs) from the COSMOS/ULTRAVista observational data ($\log(n/\rm{Mpc^3})= -2.19, -2.68$ at $z= 1.08, 2.07$). 
Indeed the models fail to match the observed SMFs at $z>1$ by over-predicting the number of galaxies below the characteristic mass ($M_*$) 
of the SMF \citep{fontanot+09, hirschmann+12, wang+12}.
This discrepancy, which gets worse at higher redshifts, arises in the central galaxy population of intermediate mass haloes, but affects also 
satellites (as centrals become satellites in a hierarchical Universe). 
We statistically correct this problem by assigning to each galaxy
a weight ($w$) which is the ratio between the predicted and the observed SMFs (i.e. number densities) at the stellar mass of the galaxy. 
For this correction, the model stellar masses are convolved with a gaussian error distribution with sigma 0.25 dex in order to match the 
uncertainties on the observed stellar masses. 
We use the the observed SMFs from \citet{muzzin+13} \footnote{We make use the single Schechter \citep{schechter+76} fit 
where the faint end slope ($\alpha$) is a free parameter. }
because the mass limit of their dataset is well below $M_*$ at the redshifts of our interest allowing a robust determination of 
the faint end slope. Moreover the data are deep enough that the SMFs are not extrapolated to the
stellar mass limit we use at $z= 1.08$, and extrapolated by only 0.5 dex at $z=2.07$. 
The weight is then used when computing the local density 
around each galaxy as presented in Section \ref{methodENV} and when statistical properties of galaxies or of their parent 
haloes are computed, unless otherwise stated.
 
To define passive galaxies we make use of the specific star formation rate ($sSFR$), i.e. the star formation rate per unit mass. 
We set the $sSFR$ limit as follows:
\begin{equation}
sSFR < b/t_z
\end{equation}
where $b$ is the birthrate parameter $b= SFR/<SFR>$ as defined by
\citet{sandage+86} and $t_z$ is the age of the universe at redshift
$z$.  Inspection of the sSFR distribution in our models revealed there
is little, if any, dependence on stellar mass, thus we use the value
proposed by \citet{franx+08}: $b=0.3$ (see Figure \ref{sSFRMs}).  The
$sSFR$ limits are $\sim 5.5\times10^{-11},~{\rm and}~
9.1\times10^{-11}~\rm yr^{-1}$ at $z= 1.08$ and $2.07$ respectively.
The quantitative results presented in this paper would slightly change
if a different limit is set.  None the less the qualitative trends are
unchanged.

\section{Quantification of environment} \label{methodENV}
For the following analysis, we use model galaxies to reproduce and test different possible measurements of the 
density on scales ranging from intra-halo to super-halo.
First of all we convert one of the comoving axes of
the simulation box into a physically motivated redshift. The centre of the box is taken to be at the 
exact redshift of the snapshot under investigation. We compute the positional offset of each 
galaxy from the centre and convert this into a redshift offset using a cosmology calculator 
\citep{wright+06}. Then, the effect of peculiar velocities is included to produce redshift distortions. 
This produces redshifts with a quality similar to high spectral resolution observations  
(hires-z hereafter). We also create two sets of less accurate redshifts.
The first one is obtained by convolving the hires-z with a gaussian error distribution with $\sigma=1000km/s$. This roughly 
corresponds to the redshift accuracy of low spectral resolution (lowres-z) surveys such as those
obtained using slitless spectroscopy on HST \citep{brammer+12}. 
Those surveys have the huge advantage of obtaining a redshift for every object in the observed field, 
reducing the selection bias of pointed spectroscopic surveys and increasing the sampling rate close to $100\%$. 
The second set is a photometric redshift (photo-z) sample 
obtained by convolving the hires-z with a gaussian error distribution\footnote{This value is 
consistent with the accuracy of photometric redshifts for galaxies as faint as our mass selection limit in 
the deep fields where a wealth of multiwavelength data is available \citep[see e.g.][]{ilbert+09, whitaker+11}.} with $\sigma = 5000km/s$. 
When the redshifts accuracy is decreased, galaxies can be scattered outside the redshift interval (which is set by the size 
of the simulation box). In those cases we assume a periodic box such that galaxies scattered beyond the maximum redshift are 
included in the front of the box and viceversa. We make use of those three samples to test the performances of our methods 
in different scenarios.

In order to obtain measurements of density we apply a method similar to the one described in \citet{wilman+10}.
We consider all the galaxies more massive than the limit set in Section \ref{Models} and we calculate the 
projected density of weighted neighbouring galaxies $\Sigma_{r_i,r_o}$ in a combination of annuli centered 
on these galaxies with inner radii $r_i$ and outer radii $r_o$. Our set of apertures ranges from 0.25 Mpc to 1.5 Mpc. 
This allows us enough flexibility to use either a single circular annulus ($r_{i}=0$) or a combination
of a circular annulus ($r_{i_1}=0$) and an outer annulus that does not overlap with the previous one ($r_{i_2} = r_{o_1}$).  
As shown by \citet{wilman+10} this method allows us to test the correlation between galaxy properties
and the density on different scales.
For an annulus described by $r_i$ and $r_o$, the projected density is computed as follows:
\begin{equation}
\Sigma_{r_i,r_o} = \frac{w_{r_i,r_o}}{\pi(r_o^2-r_i^2)}
\end{equation}
where $w_{r_i,r_o}$ is the sum of the weights of neighbouring galaxies living at a physical projected distance\footnote{
The choice of physical apertures in place of comoving is motivated by the fact that they do not depend on redshift and 
they allow for a straightforward comparison with halo sizes.} 
$r_i\leq r < r_o$ from the primary galaxy, and within a rest-frame relative velocity range $\pm dv$.
Hereafter, the density in a circle will simply be labelled as $\Sigma_{r_o}$.
The primary galaxy is not included in the sum, thus isolated galaxies have $\Sigma_{r_o} = 0$. 

The use of weights effectively changes the number of galaxies per halo or aperture 
in order to match the stellar mass function, without altering the clustering properties of haloes 
from the simulation. At intermediate to high densities this approach is sufficient to mimic the real 
universe, while at lower densities there is little dependence of the quantities we study (median 
halo mass, passive fraction) with density. In the end, the qualitative trends presented in this work 
are not different if the weights are not applied.

We use the velocity cut at $dv = 1500 km/s$ for the hires and lowres-z samples. 
This is indeed adequate for a sample with complete spectroscopic redshift coverage \citep{muldrew+12, shattow+13}, 
which will be the case for deep field surveys in the near future. Because the photometric redshifts are less accurate, we 
increase the velocity cut at $dv = 7000 km/s$, when we use this sample. This keeps the ratio between $dv$ and the 
redshift accuracy roughly constant across the three samples.

We remove from the analysis the galaxies living near the edges of the box. In spatial coordinates this affects objects closer to the 
edges than $r_o$. In redshift space we remove all objects in the first and the last $70/h$ Mpc to ensure that the cylindrical apertures
are always within the redshift limits of the sample. The final impact on the overall statistics is negligible.

In addition, we record the stellar mass rank of each galaxy with respect to its neighbours in the same 
set of cylinders. For each primary galaxy we record its rank in stellar mass with respect to the neighbouring
galaxies in the volume defined by a cylindrical aperture. If the primary galaxy is the most massive it scores rank 1, if
it is the second most massive it has rank 2 and so on. Then the cylinder is put on the following primary galaxy and the procedure is repeated. 
Because the rank in a cylindrical annulus is of little physical interest, we use regular cylinders ($r_{i}=0$). 
We show in Section \ref{RankCenSat} how the mass rank can be used to discriminate between the central 
and satellite populations. 

\section{The correlation of density with halo mass} \label{Sect_Halomassenv}
\begin{figure*}
\begin{center}
\includegraphics[width = 0.95\textwidth]{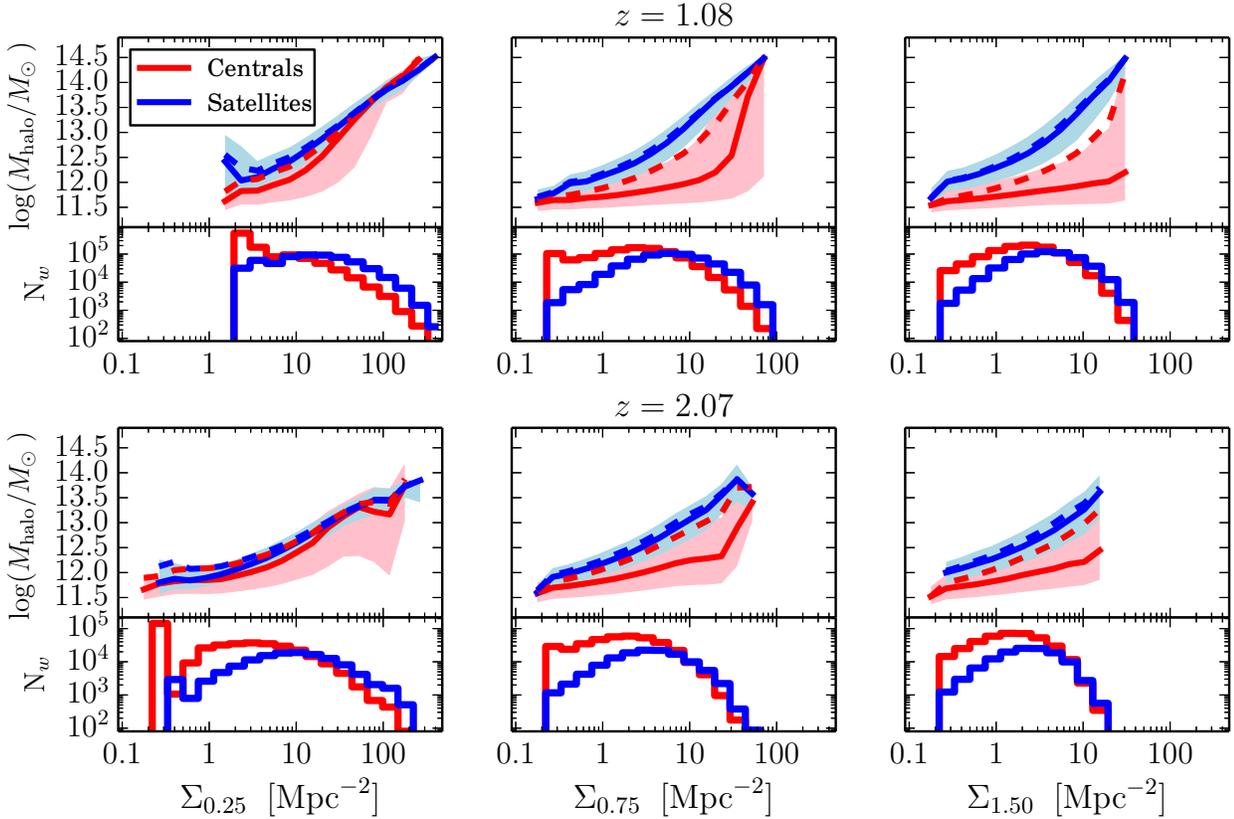}
\end{center}
\caption{Median mass of the parent halo of each galaxy as a function of density
  measured on three scales $\Sigma_{0.25}$ (left panels),
  $\Sigma_{0.75}$ (central panels), and $\Sigma_{1.50}$ (right panels)
  for central galaxies (red solid lines) and satellites (blue solid
  lines) at $z= 1.08$ (top panels), and $z= 2.07$ (bottom panels). The
  shaded areas are bounded by the 25th and 75th percentiles of the
  distributions. Dashed lines correspond to the mass of the most massive
  halo within 1 Mpc ($M_{\rm{h,1Mpc}}$, see text). The distribution of densities
  for centrals and satellites are shown in 
  the lower part of each panel.} 
\label{Halomass1Denv}
\end{figure*}

In this section we examine correlations of density measured on
  different scales with halo mass for both central and satellite
  galaxies at $z= 1.08$ and $z= 2.07$. This compliments recent work in
  the local Universe by \citet{muldrew+12}, \citet{haas+12}, and
  \citet{hirschmann+14}. 

Figure \ref{Halomass1Denv} shows the correlation of median (and 25$^{th}$
and 75$^{th}$ percentile) halo mass with density on three scales:
$\Sigma_{0.25}$ (left panels), $\Sigma_{0.75}$ (central panels), and
$\Sigma_{1.50}$ (right panels) for central galaxies (red solid lines)
and satellites (blue solid lines).
In each panel, the histograms at bottom show the weighted distribution of density for centrals 
and satellites on the same scales . The top and bottom rows refer to 
$z=1.08$ and $z=2.07$, respectively. 
The binning is logarithmic in density and galaxies with no neighbours within 
the aperture ($\Sigma_{r_o} = 0$) are included in the first bin that is
populated with objects at each scale.
From a first look at the distributions it is clear that the three different 
scales probe different ranges of density. The bigger the aperture the lower 
is the density that can be measured.  

For satellite galaxies the correlation is remarkably good at all scales and
all redshift: in this redshift range even the smallest aperture we use
(0.25 Mpc) is big enough to recover a density dependence for the
satellites. 

The halo mass dependence on density for centrals is a strong function
of the aperture size. The typical
virial radius of a $10^{13} M_\odot$ (resp. $10^{14} M_\odot$) halo is
0.30 (resp.  0.63) Mpc. As a result the 0.25 Mpc aperture probes
intra-halo scales for all reasonably massive haloes, and a good
correlation with density (which is almost indistinguishable
  from that of satellites) arises. Despite the low number counts in
  this small aperture, the correlation we find is not unexpected. It
  follows from a power law dependence of median halo mass on
  group size (defined as the number of galaxies above the stellar mass
  limit which live in the same dark matter halo). This correlation
  extends to small group sizes ($2-3$ members per group) and holds for
  centrals as well as for satellites. This happens whenever
  the aperture size does not extend well beyond the halo virial radius.
  Taking a look at the distribution of 
  density on 0.25 Mpc scale it is clear that, while a population of isolated
  centrals exists at ($\Sigma_{0.25} < 5 {\rm Mpc^{-2}}$), the higher densities 
  are also well populated by central galaxies
  It is indeed at those high densities that we find a good correlation with the group size.   

Looking at the 25$^{th}$ and 75$^{th}$ percentiles of the halo
  mass distribution at fixed density (shaded region in
  Figure~\ref{Halomass1Denv}) we notice that the trend is much tighter
  (smaller scatter) for satellite galaxies than for centrals. Indeed the 
  75$^{th}$ percentile for centrals tracks, albeit with some changes, the 
  halo mass of satellites at fixed density on all scales, while the median 
  and to a greater extent the 25$^{th}$ percentile drops to much lower halo mass
  as the scale and density increases (almost erasing any correlation
  with halo mass). In other words: there exists a significant
  population of central galaxies at high density which inhabit low mass
  haloes, and the size and relevance of this population increases to
  larger scales.

\begin{figure}
\begin{center}
\includegraphics[width = 0.95\columnwidth]{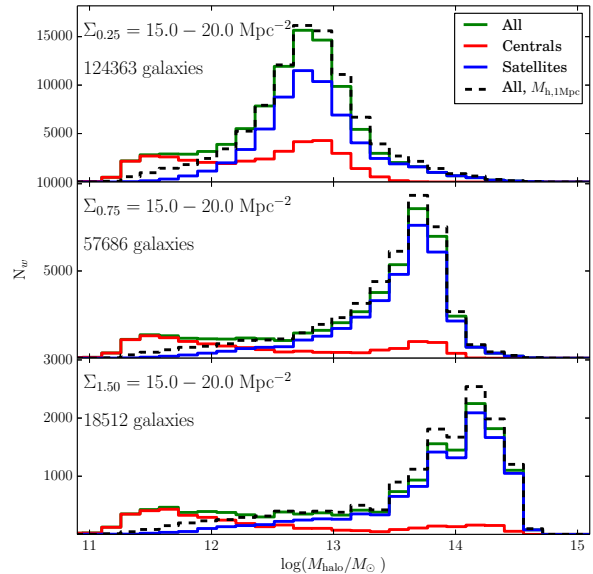}
\end{center}
\caption{Weighted distributions of parent halo mass for each galaxy in a fixed bin of density (15-20 Mpc$^{-2}$) 
on scales $\Sigma_{0.25}$ (top panel), $\Sigma_{0.75}$ (middle panel), and $\Sigma_{1.50}$ (bottom panel) at $z=1.08$. 
The dashed line shows the weighted distribution of the most massive halo mass within 1 Mpc of each galaxy (see text).} 
\label{Halomass1Dhisto}
\end{figure}

We quantify this statement in Figure \ref{Halomass1Dhisto} which shows
the halo mass distributions in a fixed bin of density (15-20
Mpc$^{-2}$) on scales 0.25 Mpc (top panel), 0.75 Mpc (middle panel),
and 1.50 Mpc (bottom panel) at $z=1.08$. This bin is chosen to probe a
fairly high density with good statistics on the three scales
although it is a more unusually large density when measured on
  larger scales.  Satellites at this density occupy a peak of
  relatively high halo mass, demonstrating the tight correlation
  between halo mass and density seen in Figure~\ref{Halomass1Denv}. A
  significant fraction of central galaxies live in the same peak,
  illustrating the relative insensitivity of a ``density within an
  aperture'' statistic to whether a galaxy is the central or a
  satellite galaxy of a massive halo. However there is also a second
  significant population of central galaxies at low halo mass: 52\%
  (61\%, 65\%)\footnote{Hereafter the first value refers to the 0.25 Mpc
    aperture while those in parenthesis refer to the 0.75 Mpc and 1.50
    Mpc apertures respectively.}. These galaxies live in small haloes
  (which is why they are usually centrals) but have a high number
  of neighbouring galaxies which live within the
  cylindrical aperture used to measure density -- a number which can
  only increase with the aperture size.
 This phenomenon is important when a significant fraction of the
  neighbouring galaxies used to trace density live outside the galaxy's
  own host halo -- i.e. when the aperture scale is larger than the
  virial radius of the halo. This explains why the scatter becomes
  small when densities are measured on a 0.25 Mpc
  scale, while on the 0.75 Mpc scale (at $z=1.08$) the median shoots up
  at the very highest densities -- such densities are most often
  obtained at the centres of massive haloes which have virial radii
  close to 0.75 Mpc. However, in all other regimes, the low halo mass
  population is the dominant one for central galaxies at intermediate
  to high density.

Our goal is to calibrate the environment of galaxies using
  measurements of density: Figure~\ref{Halomass1Dhisto} shows that this
  is a degenerate problem where we have only a single measurement of
  density within an aperture. In Appendix \ref{sec:multiscales} we
  examine how the combination of two scales can break this degeneracy,
  while in Section \ref{passivefracvsenv} we ignore the
  second peak by excluding low (stellar) mass galaxies at high density.
  Trends driven by the population of low halo (or stellar) mass
  centrals at high density are difficult to interpret because they can
  have actual physical association with the nearby massive halo to
  which they have (at the current snapshot) not been assigned.  Using
  an N-body simulation and accurately tracing the trajectories of
  ejected satellites, \citet{wetzel+13} have shown that infalling
  galaxies can pass through a massive halo on radial orbits and emerge
  out the other side, where they extend out to 2.5 times the virial
  radius of the halo they crossed. They compose 40\% of all central
  galaxies out to this radius and their evolution is likely to have
  been influenced by satellite-specific processes. This ``backsplash''
  population has also been investigated by \citet{mamon+04, balogh+00, ludlow+09}
  and \citet{bahe+13}, as well as \citet{hirschmann+14} who also find 
  that such additional processing of low mass, high density galaxies is 
  necessary to explain the density dependence of the passive fraction of central
  galaxies at low redshift. Thus an accurate accounting for this
  population is essential. 

  To examine the actual real-space proximity of galaxies
  to massive haloes, we consider the most massive halo within a sphere of 1
  Mpc ($M_{\rm{h,1Mpc}}$). In Figure \ref{Halomass1Dhisto} we show the
  distribution of $M_{\rm{h,1Mpc}}$ (black, dashed line). This does not
  completely exclude the low halo mass population, but reduces it
  substantially. The fraction of centrals with $M_{\rm{h,1Mpc}} <
  10^{12.5} M_\odot$ is 37\% (32\%, 30\%), far fewer than where the
  host halo mass is used. Most remaining such galaxies are in this
  high-density bin due to redshift space projection. However this tells
  us that almost half of low halo mass central galaxies at this density
  are within 1 Mpc of a massive halo and many may have already
  passed through that halo. Thus for a clean selection of central
  galaxies which have not suffered such effects, it seems sensible to
  exclude those at high density. In Figure \ref{Halomass1Denv} we
  overplot the median relation of density with $M_{\rm{h,1Mpc}}$
  (dashed line). The reduced peak at low mass means that this more
  closely follows the 75$^{th}$ percentile of halo mass, and approaches
  that of satellites (for which the low halo mass peak is negligible).

\begin{figure}
\begin{center}
\includegraphics[width = 0.95\columnwidth]{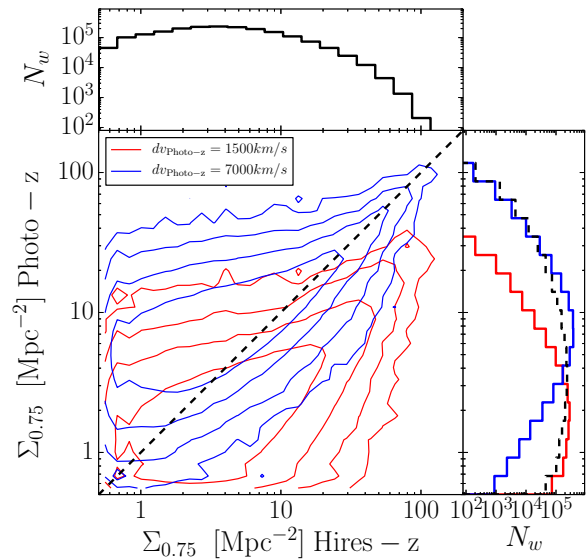}
\end{center}
\caption{Main panel: Density on the 0.75 Mpc scale for the photo-z sample and different velocity cuts ($dv=1500 km/s$ red 
and $dv=7000 km/s$ blue) as a function of density in the same aperture (and $dv=1500 km/s$) for the hires-z sample. The contours are 
logarithmically spaced with the outermost contours at 25 objects per bin and the innermost at $10^4$ objects per bin.
Upper panel: weighted distribution of density for the hires-z sample. Right panel: weighted distributions of density 
for the photo-z sample with the same velocity cuts as above. The distribution of density for the hires-z sample (dashed black) is 
repeated here for comparison.} 
\label{Densitycomp.}
\end{figure}

  In order to test the effect of redshift accuracy, we repeat the analysis
  on the lowres-z sample.  Our results show that none of the trends presented in this
  section notably change, the main effect being a smoothing of the highest density 
  peaks, slightly reducing the median halo mass. Conversely, when photometric redshifts
  are used we notice two effects. First, the median halo mass at fixed density is reduced
  both for centrals and satellites on all scales. The 0.75 and 1.50 Mpc scales
  are the most affected. The median halo mass for centrals is
  constant with density at $10^{12} M_\odot$. For satellites this is 
  0.5 dex higher and increases with density only at $\Sigma > 10 {\rm Mpc^{-2}}$.
  On the 0.25 Mpc scale a good correlation of median halo mass with 
  density still exists, but with a larger scatter both for centrals and satellites.
  The second effect we find is in the distributions of density, which are narrower 
  than in the hires-z case, reducing the density dynamic range. In Figure \ref{Densitycomp.}
  we compare the density in the hires-z sample to those in the photo-z sample for the 
  0.75 Mpc aperture. The black dashed line marks the 1:1 relation. 
  We show two velocity cuts for the photo-z sample: $dv=1500 km/s$ (red) 
  and $dv=7000 km/s$ (blue). In the upper panel the distribution of densities is plotted
  for the hires-z sample and in the right panel we show the density distributions for
  the two apertures used for the photo-z sample. 
  Both from the contours and the histograms it is evident that $dv=1500 km/s$ misses 
  many objects in the most dense regions, while with a larger depth $dv=7000 km/s$ we 
  re-incorporate those galaxies obtaining a good correlation with the hires-z measurements.
  Moreover, using photo-z, the low densities are devoid of galaxies which end up at 
  intermediate densities. Also this effect is less severe for the 0.25 Mpc aperture. 
  Thanks to the large statistics, we can probe a wide range of density in the 
  models, however this is reduced when considering the number of objects in a real survey.

Finally we test if the distributions shown in figure
\ref{Halomass1Dhisto} would change at the low halo-mass end due to the
resolution of the N-body simulation.  The history of galaxies whose
parent haloes are below $10^{12} M_\odot$ cannot be traced accurately
along the halo merger trees, thus their physical properties might be
inaccurate. We perform the same exercise using the G13 model applied to
the Millennium-II simulation scaled to WMAP7 parameters following
\citet{angulo+10}.  This simulation has a smaller cosmological
volume but a particle resolution about 100 times better. The
distribution of halo masses for centrals and satellites below $10^{12}
M_\odot$ is unchanged, probably thanks to our conservative limit in
stellar mass.

\section{Mass rank as a method to disentangle centrals and satellites} \label{RankCenSat}

\begin{figure*}
\begin{center}
\includegraphics[width = 0.90\textwidth]{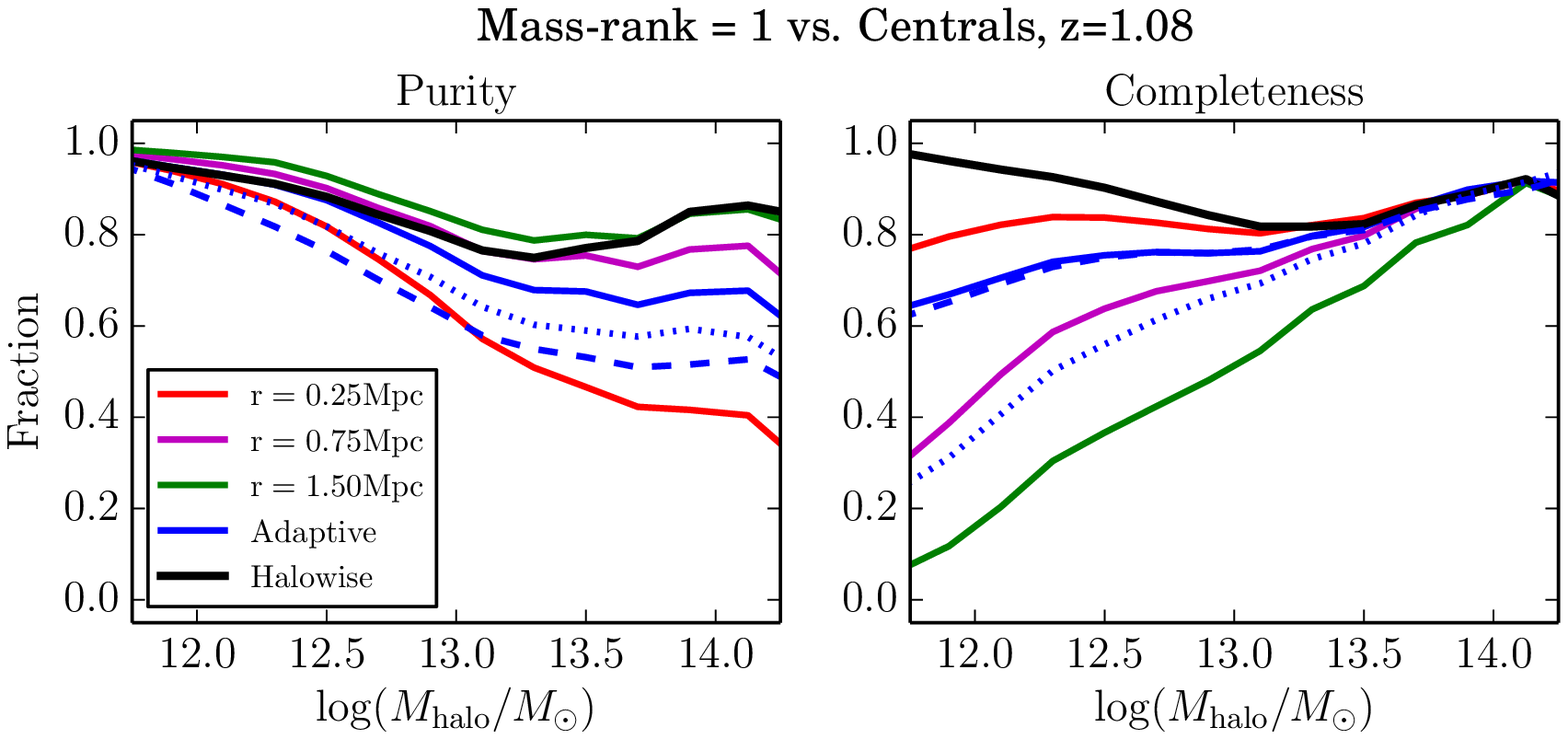}
\includegraphics[width = 0.90\textwidth]{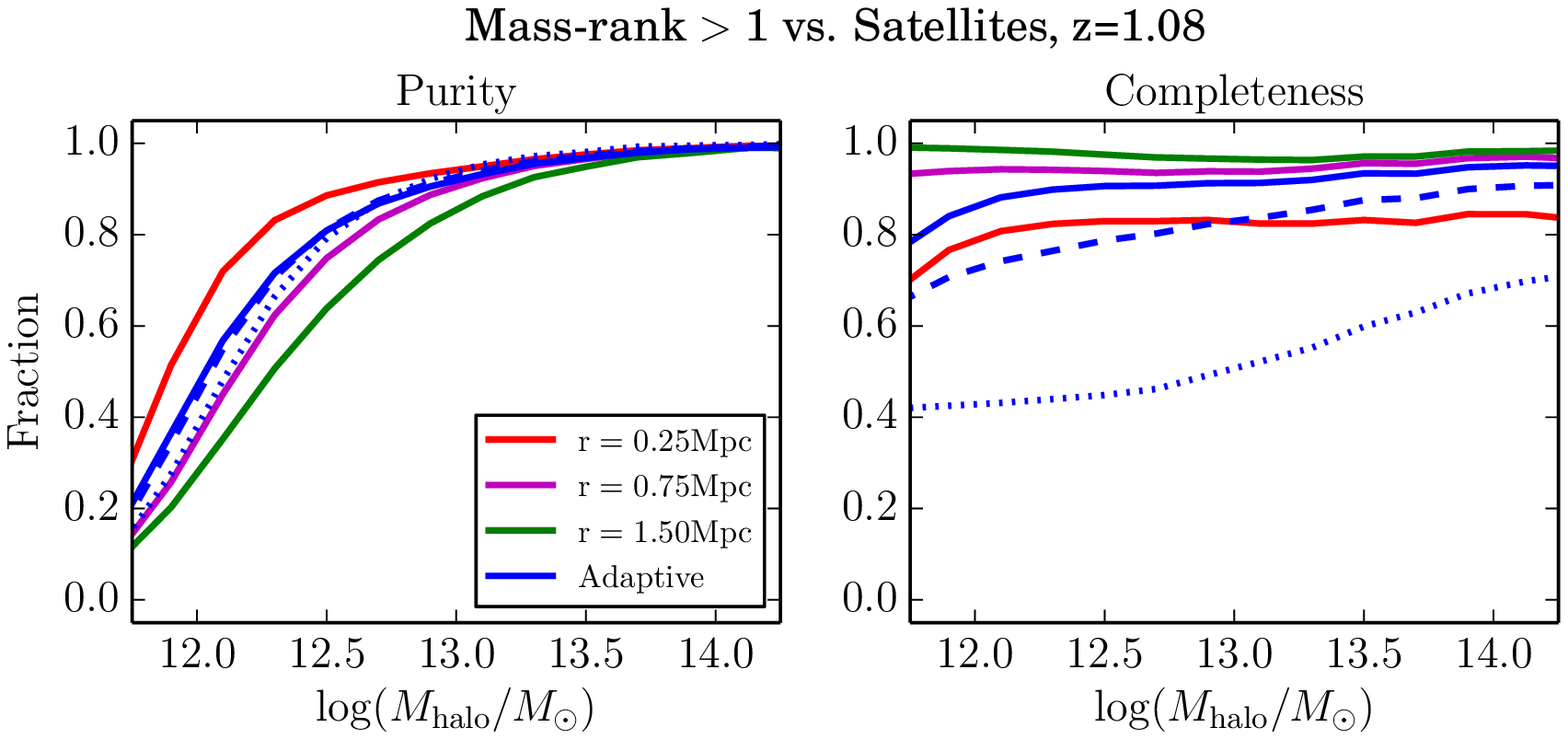}
\includegraphics[width = 0.90\textwidth]{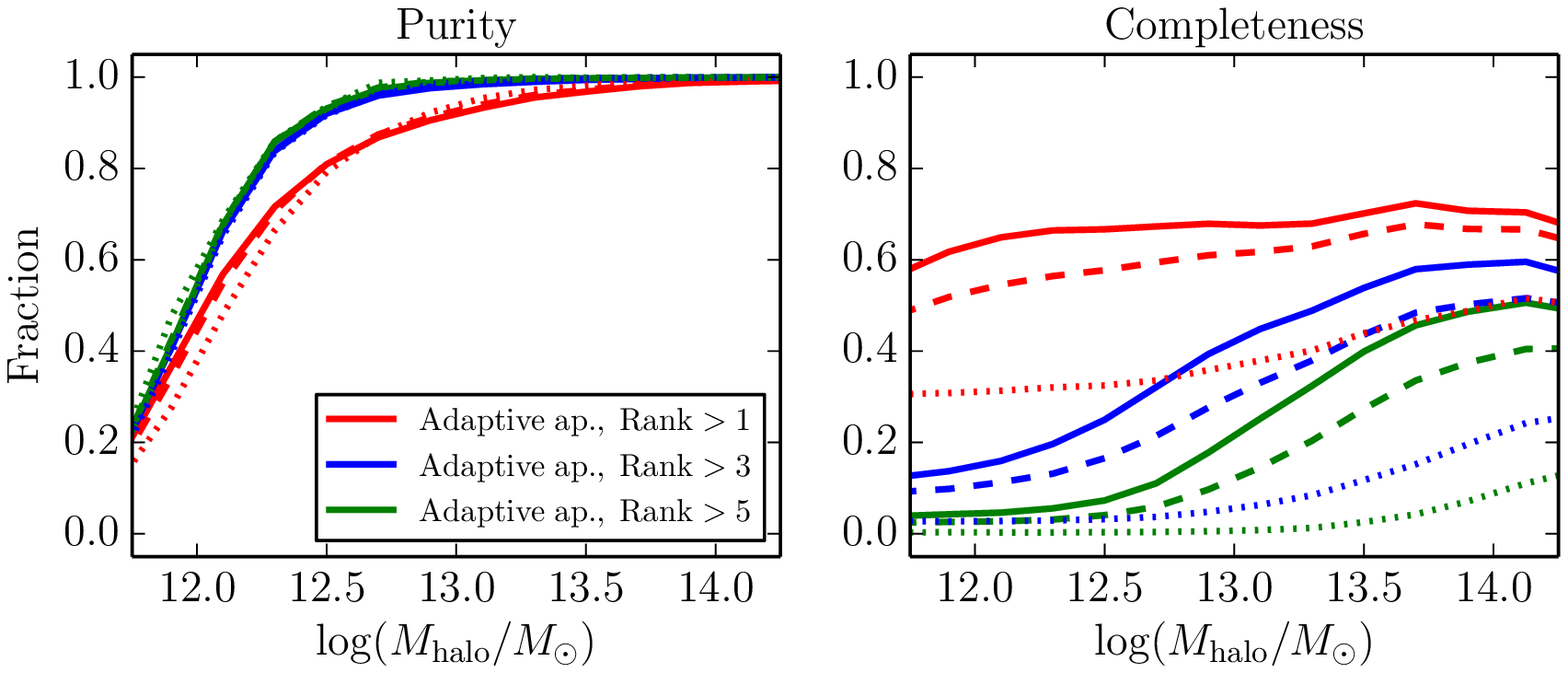}
\end{center}
\caption{Top panels: purity and completeness for the selection of central galaxies based on stellar mass rank = 1 as a function of the halo mass. The rank 
is computed in various apertures ranging from 0.25 to 1.50 Mpc for the $z=1.08$ bin. Solid lines refer to the hires-z, dashed lines to the lowres-z 
and dotted lines to the photo-z samples. Middle panels: same as before but for satellite galaxies selected by having stellar mass rank $> 1$. Bottom panels: 
comparison between different rank criteria for the selection of satellite galaxies using the adaptive aperture. } 
\label{SelRank}
\end{figure*}

It is also critical to describe whether a galaxy dominates its
halo (and the local gravitational field) or if it is instead orbiting
within a deeper potential well. This can be modelled (and is in SAMs)
assuming a dichotomy between central and satellite galaxies.  
Central galaxies accrete gas by
cooling, and merge with their satellites.  In contrast, satellite
galaxies orbit within the gravitational field, and move with respect to
the intra-halo gas, thus experiencing tidal interaction and stripping
effects.  This has been both directly observed in nearby clusters
\citep[see e.g.][]{ boselli+08, yagi+10, fossati+12} and
indirectly witnessed from statistical studies \citep{balogh+04}.
Moreover, it has been claimed that, while the properties of
  galaxies are shaped by intrinsic parameters, (e.g. stellar
  mass \citealt{peng+10, kovac+13}, see however \citealt{delucia+12}),
  satellites' properties are also influenced by their environment
  \citep{peng+12, woo+13}.  A reliable method to separate centrals
from satellites in observations is therefore crucial. 

\subsection{Identification of central galaxies} \label{selcentral}

Central galaxies are usually the most massive galaxy in 
their halo (but not always, see \citealt{skibba+11}). This follows from
the fact that the other -- satellite -- galaxies in the halo were
formed at the centre of less massive progenitor haloes. Therefore a
sample of central galaxies can be identified by assuming that the most
massive galaxy in a halo is also its central galaxy
\citep[e.g.][]{yang+08}.

We compute the rank in stellar mass of each galaxy in several circular
apertures. We examine both fixed radius apertures, and a radius
that depends on stellar mass (accessible from observations). 
This approach resembles the Counts-in-Cylinders method (\citealt{reid+09}, see 
also \citealt{trinh+13}). However, it is worth stressing that previous 
applications of this method were focused on different science goals 
(e.g. the identification of 2 member groups in a specific survey). 
We present here a detailed analysis of how much a population of galaxies 
whose stellar mass rank is 1 compares to galaxies identified as centrals
by the algorithms used in the models (see sec. \ref{Models}).

We define two parameters to quantify the overlap between the two
populations. The purity ($P$) is the number of centrals which are
correctly identified over the number of selected galaxies; and the
completeness ($C$) is the number of identified centrals over the total
number of central galaxies.

In this section (and in Sections \ref{selsatellite} and
\ref{PandCsellim}) the use of the weights for each galaxy has a
negligible impact on both the qualitative and the quantitative results.
The method is insensitive to the overestimation of the number of low
mass galaxies because the galaxies which compete to be the most massive
in an aperture have similar stellar masses, thus the same weight.
Therefore we do not use the weights in this Section.

In the top panels of Figure \ref{SelRank} the purity and the
completeness of identified centrals are plotted as a function of halo
mass. Solid coloured lines correspond different apertures for the hires-z sample
ranging from 0.25 Mpc to 1.5 Mpc. Ideally one would like to maximise both $P$ and
$C$ but a trade-off must be found. In table \ref{PandCcent} the
performances of different methods are given. Purity and completeness
are given for haloes above and below $10^{13} M_\odot$ and for the
complete sample.

We start by identifying the ``halo-wise'' mass rank of each galaxy
by ranking in stellar mass all the galaxies belonging to the same
halo. We obtain the black solid line in the top panels of
figure \ref{SelRank}.
The purity and completeness of this sample
describes the ideal overlap between mass rank 1 and central galaxies.  
The completeness is limited by the fact that the
most massive galaxy is not always the central of its host halo.  This
is true especially at halo masses around $10^{13} M_\odot$ where this
incompleteness reaches 20\%. 
Due to the scatter in the
stellar-mass halo-mass relation for central galaxies, haloes of masses
$\sim 10^{12-13}M_\odot$ can all host galaxies of equivalent stellar
mass which means that even in minor halo mergers (down to a mass
ratio of $\sim 1:10$) a more massive galaxy can be supplied by the
less massive halo.  
It will then become a satellite more massive than the central of the final halo.  
The purity and completeness estimates are always computed relative to the
population of ``true'' central galaxies in the model -- as such the
halo-wise values provide an upper limit on completeness, and
an``optimal'' purity based on the assumption that we perfectly know the
content of each halo. However the most massive galaxy in any halo
corresponds to a significant local potential, and as such one could
also define a purity and completeness relative to this population.
Such estimates of purity and completeness would clearly be significantly
higher than those defined here (relative to the central population). 

Looking at Figure \ref{SelRank}, it is clear that fixed apertures (red,
magenta, and green solid lines) struggle to balance the requirements 
for both high purity and completeness.
The general trend is for an increasing purity and decreasing
completeness as the aperture size is increased. However, for small
apertures (0.25 Mpc) $P$ goes down at the high halo mass end because
the aperture covers only a fraction of the halo; if they are too large
then $C$ goes down in smaller haloes because one aperture covers
multiple haloes.

From this evidence and based on the idea that a good correlation exists between 
stellar mass and halo virial radius (which is an analytic function of halo mass) for centrals, we 
define an ``adaptive'' aperture as follows:
\begin{equation}
r = \rm{min}(0.75,n\times10^{(\alpha \log M_{*}+\beta)})~\rm{[Mpc]}
\end{equation}  
where $M_*$ is the stellar mass of the galaxy, $n$ is a multiplicative factor, and $\alpha$ and $\beta$ are 
the parameters which describe the dependence of virial radius on stellar mass for centrals in the models
\footnote{The $\alpha$ and $\beta$ parameters are obtained by fitting a linear relation ($r=0.89$) in log-log space
between the virial radius and the stellar mass for the central galaxies in the models.}. 
From the models we get $\alpha=0.25$, $\beta=-3.40$ at all redshifts and after extensive testing we define 
$n=3$ in order to avoid very small apertures that would decrease the purity. In order to limit the size of 
the aperture to a scale that entirely covers the most massive haloes without extending beyond, 
we limit the aperture to a radius of 0.75 Mpc. From Figure \ref{SelRank} and from the values in table 
\ref{PandCcent} it is evident that this is a great improvement for $C$ at low halo masses relative to the 
fixed 0.75 Mpc aperture, while we lose 8\% in purity at high halo masses. This is because we are using a small 
aperture for low mass galaxies (0.2 Mpc for $M_*=10^{9.5} M_\odot$). Since some of them are satellites living in 
massive haloes, the apertures we use are too small to encompass the entire halo and those satellites 
can get high mass rankings reducing $P$.

\begin{table}
\begin{center}
\begin{tabular}{c c c c c c c }
           & \multicolumn{2}{|c|}{$\log M_{\rm h} \leq 13 $} & \multicolumn{2}{|c|}{$\log M_{\rm h} > 13$} & \multicolumn{2}{|c|}{All} \\
Aperture   & $P$ & $C$ & $P$ & $C$ & $P$ & $C$ \\
\hline
 & \multicolumn{5}{|c|}{$z=1.08$} & \\
0.25 Mpc    & 0.92 & 0.82 & 0.52 & 0.87 & 0.88 & 0.82 \\ 
0.75 Mpc    & 0.94 & 0.40 & 0.75 & 0.78 & 0.92 & 0.42 \\
1.50 Mpc    & 0.94 & 0.17 & 0.80 & 0.64 & 0.91 & 0.19 \\
Adaptive    & 0.94 & 0.68 & 0.75 & 0.78 & 0.92 & 0.69 \\
\hline							
 & \multicolumn{5}{|c|}{$z=2.07$} & \\
0.25 Mpc   & 0.93 & 0.79 & 0.76 & 0.91 & 0.92 & 0.79 \\
0.75 Mpc   & 0.95 & 0.38 & 0.86 & 0.83 & 0.94 & 0.40 \\
1.50 Mpc   & 0.95 & 0.18 & 0.88 & 0.71 & 0.94 & 0.20 \\
Adaptive   & 0.94 & 0.64 & 0.84 & 0.86 & 0.94 & 0.65 \\
\end{tabular}
\caption{Purity and completeness for the identification of central galaxies using stellar mass rank = 1 in four different apertures 
for haloes below and above $M_{\rm halo} = 10^{13} M_\odot$, and for the full hires-z sample at $z=1.08$ and $z=2.07$.} 
\label{PandCcent}
\end{center}
\end{table}

\subsection{Identification of satellite galaxies} \label{selsatellite}
\begin{table}
\begin{center}
\begin{tabular}{c c c c c c c }
           & \multicolumn{2}{|c|}{$\log M_{\rm h} \leq 13 $} & \multicolumn{2}{|c|}{$\log M_{\rm h} > 13$} & \multicolumn{2}{|c|}{All} \\
Aperture   & $P$ & $C$ & $P$ & $C$ & $P$ & $C$ \\
\hline
 & \multicolumn{5}{|c|}{$z=1.08$} & \\
0.25 Mpc   & 0.58 & 0.78 & 0.97 & 0.79 & 0.70 & 0.78 \\
0.75 Mpc   & 0.33 & 0.92 & 0.96 & 0.93 & 0.45 & 0.93 \\
1.50 Mpc   & 0.27 & 0.97 & 0.91 & 0.96 & 0.38 & 0.96 \\
Adaptive   & 0.46 & 0.86 & 0.95 & 0.91 & 0.59 & 0.88 \\
\hline						       
 & \multicolumn{5}{|c|}{$z=2.07$} & \\
0.25 Mpc   & 0.49 & 0.79 & 0.93 & 0.83 & 0.56 & 0.80 \\
0.75 Mpc   & 0.28 & 0.92 & 0.90 & 0.92 & 0.33 & 0.93 \\
1.50 Mpc   & 0.24 & 0.96 & 0.85 & 0.94 & 0.29 & 0.95 \\
Adaptive   & 0.37 & 0.84 & 0.92 & 0.91 & 0.44 & 0.84 \\
\end{tabular}					       
\caption{Purity and completeness for the identification of satellite galaxies using stellar mass rank $>1$ in four different apertures 
for haloes below and above $M_{\rm halo} = 10^{13} M_\odot$, and for the full hires-z sample at $z=1.08$ and $z=2.07$. }
\label{PandCsat}
\end{center}
\end{table}

The mass rank method can also be used to identify satellite galaxies,
under the assumption that satellites are less massive than the central
galaxy of their own halo, i.e. mass rank $>1$.  In the middle panels of
Figure \ref{SelRank} are plotted the purity and completeness (defined
as in Section \ref{selcentral}) of our identified satellite galaxies as
a function of halo mass. The purity is limited -  
but only by ~5\% - by satellites that are more massive than the central 
galaxy of their host halo. The purity quickly drops at halo masses below
$10^{12.5} M_\odot$, with little dependence on the aperture size.  In
Table \ref{PandCsat} we present the values of $P$ and $C$ for satellite
galaxies living in haloes less and more massive than $10^{13} M_\odot$,
and for the complete sample. It is clear that the overlap between
mass rank $>1$ and satellites is strong among massive haloes, while in
less massive haloes about half of the galaxies with mass rank $>1$ are
centrals which by chance have one (or more) more massive neighbours
within the aperture but not within the same halo. It is hard to
define a satellite in this halo mass regime, especially where the
stellar mass of central galaxies approaches the mass limit: the mass
of any satellites included in the sample must therefore be close to
that of their central. This halo mass regime consists primarily of isolated
galaxies and loose proto-groups (namely pairs or triplets).
Completeness in contrast is remarkably high over the full range of halo
masses, illustrating that our method is effective in picking up a
complete population of satellites.
\begin{figure*}
\begin{center}
\includegraphics[width = 0.90\textwidth]{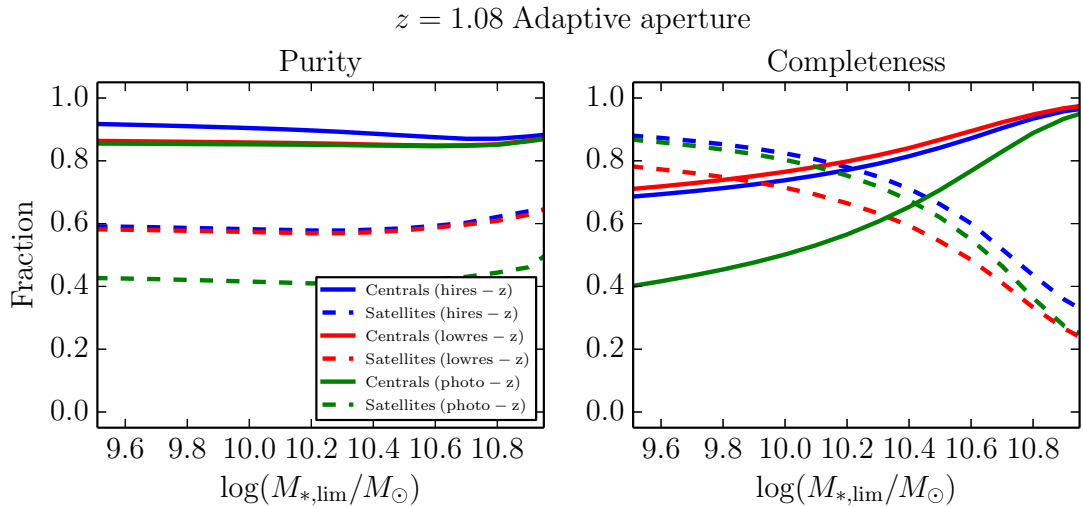}
\end{center}
\caption{Purity and completeness for the identification of central (solid lines) and satellite (dashed lines) galaxies using stellar mass rank = 1 ($>1$) 
in the adaptive aperture at $z=1.08$ for the hires-z sample (blue lines), for the lowres-z sample (red lines) and for the photo-z sample 
(green lines).}  
\label{RankMlim}
\end{figure*}

In the bottom panels of Figure \ref{SelRank} we explore the dependence of $P$ and $C$ on the stellar mass rank we required to identify a 
satellite within the adaptive aperture. Setting two more conservative limits (rank $>3$ blue solid and $>5$ green solid) we achieve a small 
improvement in the purity at low halo masses but with a dramatic degradation of the completeness. It is clear that with rank to be $>5$ ($>3$) we are 
implicitly selecting only galaxies that have at least five (three) companions within the aperture, all of them more massive. 
Moreover, especially at low/intermediate halo masses, the number of galaxies  within a halo is often smaller than five (three).
Thus many real satellites are missed. 

Based all this evidence our preferred
method to identify central (satellite) galaxies is to require a stellar
mass rank $=1$ ($>1$) in the adaptive aperture. The purity and
completeness values for this aperture are $P=0.93$ and $C=0.67$
for centrals and $P=0.60$  $C=0.90$ for satellites in the
hires-z sample at $z=1.08$.  The use of the same aperture for both
types has the advantage of making the two selections mutually
exclusive.  If two different apertures are used for selecting centrals
and satellites, the fraction of interlopers has to be taken into
account as one galaxy might be the most massive in one aperture but not
in the other one.  Finally it is worth stressing that both the
purity and the completeness of the full samples of centrals and
satellites are completely independent of the measured density and
any effort to calibrate halo mass. The key parameters
are the scale over which the rank is computed and the assumption that
central galaxies are typically more massive than nearby satellites.
This assumption holds well in the G13 models and is typically assumed
to be true in group reconstruction using observational data
\citep[e.g.][]{yang+07}.

\subsection{Dependence of Purity and Completeness on the stellar mass selection limit} \label{PandCsellim}

To avoid model resolution biases in our definition of the environment,
we probe down to a constant mass limit of $M_* = 10^{9.5} M_\odot$.
This is deeper than most observational surveys at these redshifts. 
To examine more realistic survey depths, we now test
how our selection methods perform as a function of the stellar mass
limit $M_{*,\rm{lim}}$. In Figure \ref{RankMlim} we show $P$ and $C$
for the identification of both centrals and satellites as a function of
$M_{*,\rm{lim}}$ using the adaptive aperture.  
The purity is almost unaffected by the selection 
limit while, as expected, completeness is. 
Regarding the centrals, the overall purity is about
$90 \%$ and does not depend on the aperture nor on the mass limit.
Conversely $C$ is a strong function of the minimum stellar mass.
When low mass galaxies are removed, the sample of centrals with
mass rank 1 becomes more and more complete. The completeness 
never reaches unity because, as discussed before, there are 
satellites which are more massive than the centrals of their own haloes.

As already discussed, the purity for the satellites is about 60\% for
all stellar mass limits. This happens because, close to the
stellar mass limit and within 0.75 Mpc, two-halo pairs containing
similar mass centrals are just as common as pairs of similar mass
galaxies within one halo. For this reason there is an improvement
using the adaptive aperture as it is smaller than 0.75 Mpc at low stellar
masses and this helps in limiting the contamination from centrals. The
completeness shows a well defined decreasing trend at increasing mass
limits because the higher $M_{*,\rm{lim}}$, the higher the chance that
a satellite is more massive than the central of its halo. This,
combined with the low number of massive satellites, makes the fraction
drop below 40\% at $M_{*,\rm{lim}}> 10^{10.8} M_\odot$.

These results stand even when the stellar masses are convolved 
with a gaussian random error to mimic observational uncertainties. We tested
the effect of errors up to a relatively large value of 0.5dex.
The purity for centrals and the completeness in the identification of satellites decrease
by less than 5\%, while the purity for satellites and the completeness of 
centrals decrease by 10\%. Moreover,
the trends as a function of the mass uncertainty are smooth and the values given here
are to be considered upper limits. 

\subsection{Dependence on Redshift Accuracy}

The trends discussed so far have been drawn using the hires-z sample. In this 
section we analyse how they change using less accurate redshifts. Dashed lines 
in Figure \ref{SelRank} are for the lowres-z sample and dotted lines are for the 
photo-z sample using only the adaptive aperture. The general trends apply 
to the fixed apertures as well. 

Concerning central galaxies, the purity is decreased by 
$\sim 10-15\%$, due to the smoothing of the density field 
introduced by lowres-z and photo-z. This happens because massive centrals 
are projected outside the cylinder of less massive satellites (the latters then scoring 
a mass rank 1, thus reducing the purity of the selection ). The completeness on the other hand 
is not affected in the lowres-z sample because massive centrals are still identified as 
the most massive galaxies in their cylinders. The larger velocity cut we use for the photo-z 
sample has a positive effect on the purity but decreases the completeness. This is not unexpected
because the larger velocity cut increases the volume of the cylinder where we compute the mass rank,
therefore the final effect is similar to increasing the size of the aperture.

In the case of satellites the purity is not affected by the smoothing in the redshift 
direction because the population of rank $> 1$ galaxies is still mainly composed of satellites. 
Conversely the completeness is decreased by $\sim 10\%$, and $\sim 40\%$  for the lowres-z 
and photo-z respectively. As photo-z are much less accurate than spectroscopic
redshifts it is worth noting why the performance of the method is still reasonably good.
When the mass rank based method identifies a galaxy as a satellite, it does not mean we know
it to be a satellite of a specific halo. Therefore the use of photo-z can project both the 
``true'' central and one (or more) satellites outside their original halo. When this happens, 
the central/satellite status is preserved and a satellite galaxy is now identified as central 
in the original halo. If, as it is more likely, only satellites are projected outside the 
halo, the most massive is identified as a central, thus reducing the satellites' completeness.

When the less accurate redshifts are used, we find no major 
impact on the trends of $P$ and $C$ on the stellar mass limit. The strongest
effect is found on the completeness of the identification of satellites which follows 
the same trend described above.

\subsubsection{Dependence of Purity and Completeness on the spectroscopic sampling rate} \label{PandCsampl}

\begin{figure*}
\begin{center}
\includegraphics[width = 0.90\textwidth]{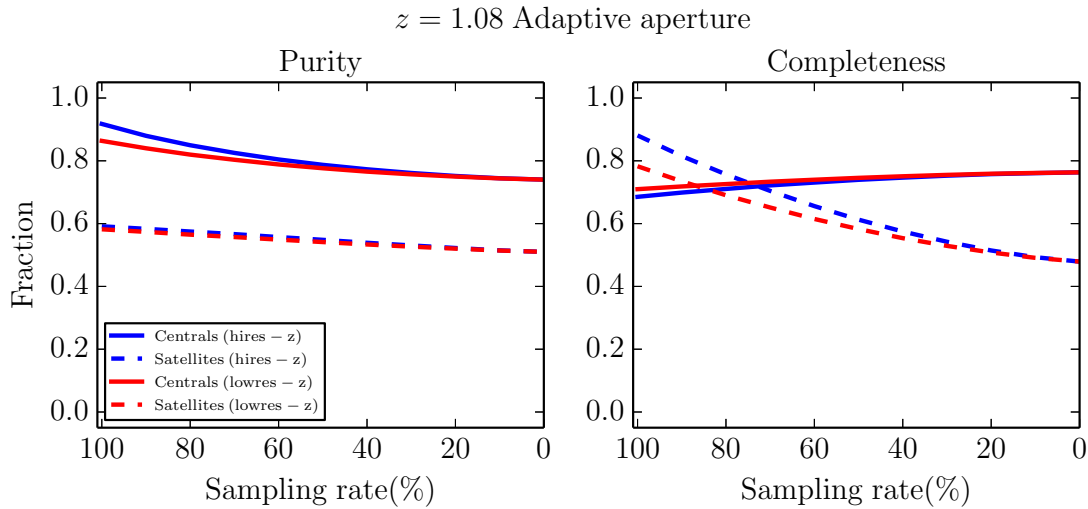}
\end{center}
\caption{Purity and completeness for the identification of central (solid lines) and satellite (dashed lines) galaxies 
using stellar mass rank = 1 ($>1$) in the adaptive aperture for incomplete hires-z (blue lines) and lowres-z (red lines) 
sampling at $z=1.08$.}  
\label{RankSampl}
\end{figure*}

Here we examine the performances of our method in the case of variable
sampling rate. Our approach is to progressively reduce the spectroscopic sampling rate from 
$100\%$ to $0$ in steps of $10\%$ by randomly replacing the hires-z (or lowres-z) with photo-z. 
Figure \ref{RankSampl} shows the purity and completeness in the adaptive aperture at each sampling rate.
The main conclusions have been discussed above using pure spectroscopic (hi- and low-res) redshifts and photo-z.
Here we only note that the decline in purity is roughly linear as a function of the sampling rate for both centrals
and satellites and that the completeness for satellites decreases more significantly from full sampling rate to 
$50\%$ rather than from this value to pure photo-z.

\begin{figure*}
\begin{center}
\includegraphics[width = 0.95\textwidth]{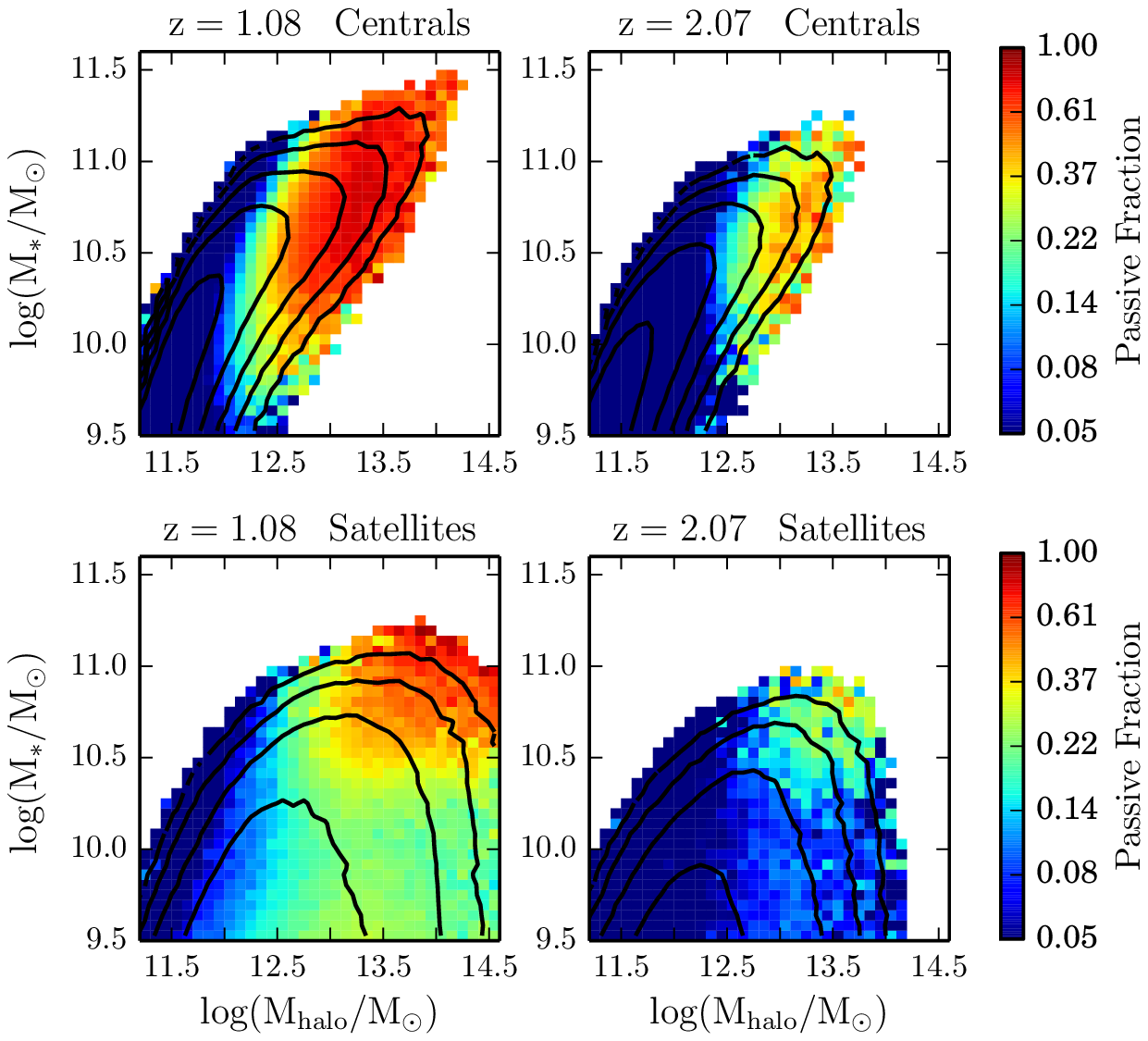}
\end{center}
\caption{Fraction of passive galaxies (as defined in Section \ref{ModelSEL}) as a function of halo mass and stellar mass for centrals 
(top panels) and satellites (bottom panels).  Redshift increases from $z= 1.08$ (left panels) to  $z=2.07$ (right panels). The contours are drawn from the density
of galaxies in the parameters space and are log-spaced with the outermost contour at 25 objects per bin and the innermost at $10^4$ objects for 
centrals and $10^{3.4}$ for satellites.} 
\label{PassfracMhMs}
\end{figure*}

Several other works have attempted a selection of central and satellite galaxies in real spectroscopic surveys. Among them \citet{knobel+12}
used a probability based method to assign a binary central/satellite classification to the I-band flux limited sample in the zCOSMOS survey.
The spectroscopic sampling rate is about $50\%$ in the redshift range $0.1-0.8$. Despite the different identification method and the broad 
redshift range (which hampers the definition of a fixed stellar mass limit), their final results ($P=0.81~C=0.89$ and $P=0.62~C=0.45$ for 
centrals and satellites respectively) agree with our results at the same sampling rate within $\sim 5\%$ for the purity and $\sim 15\%$ for 
the completeness.

In conclusion, it is remarkable how the purity and the completeness both centrals and satellites are not strongly 
affected by low spectroscopic sampling rates nor by the survey detection threshold, this proving the robustness of the method 
and its usefulness in surveys with different designs.

\section{Relation between environment and passive fraction} \label{passivefracvsenv}

In this section, we examine if and how well the environmental trends
predicted by the models can be recovered using quantities accessible
from observations, e.g. density and mass rank. We focus our attention
on a single physical quantity: the fraction of passive galaxies. \citet{peng+10,
peng+12} and \citet{woo+13} have shown that the
passive fraction of satellite galaxies correlates strongly with a
measurement of local density, labelled ``environment'', while for
centrals it is a function of their stellar mass \citep{peng+10,
peng+12} and halo mass \citep{woo+13}. 

\subsection{The growth of a passive population in the models}

First of all we investigate how the fraction of passive galaxies
depends on stellar mass, halo mass and central/satellite status.  The
former basically describes the integrated star formation and merger
history of a galaxy, while the latter two are strongly correlated
  to the regulation of its star formation in the models.  Figure
\ref{PassfracMhMs} shows the passive fraction in the $M_*-M_{\rm halo}$
space for centrals (top panels) and satellites (bottom panels), at
  redshifts 1.08 (left) and 2.07 (right). The contours are drawn
from the density of galaxies in the parameters space and are
logarithmically spaced with the outermost contour at 25 objects 
per bin and the innermost at $10^4$ objects for centrals and $10^{3.4}$ 
for satellites.  Each bin has to contain at least 10
objects for the passive fraction to be computed, and the color coding
is scaled logarithmically.

A common feature across the redshift bins is that centrals and
satellites populate different regions of the parameters space. The
centrals populate a sequence where the stellar mass
linearly increases with halo mass (in log-log space).  In contrast the
satellites form a cloud that spans all halo masses such that $M_{\rm
halo}^{\rm sat}(M_*) \geq M_{\rm halo}^{\rm cen}(M_*)$.  
  
At halo masses above $10^{12.5} M_\odot$ a passive population of centrals 
starts to appear at $z\sim2$, becoming dominant as the redshift 
decreases to 1. Those passive centrals move to higher halo masses without 
increasing their
stellar mass enough to stay on the relation defined by the star forming
centrals. The passive fraction increases to about $80\%$ for centrals
in haloes more massive than $10^{13} M_\odot$.  \citet{wilman+13},
using \citet{wang+08} models (an early version of G13) at $z=0$,
showed that a strong correlation between bulge growth and passive
fraction exists for massive centrals in SAMs. The physical reason is
that for those galaxies the cold gas reservoir is exhausted by a merger
induced starburst and further cooling of the gas is prevented by the
strong radio-mode AGN feedback. Although these trends do not perfectly
reflect the observed data, they are qualitatively similar
to those shown by \citet{kimm+09} at $z=0$ for different published
SAMs. In their work the model by \citet[][another early version of G13]{delucia+07} 
appears to be the closest to the observational constraints.

Also the passive fraction of satellite galaxies shows significant evolution.
As time goes by (and redshift decreases) the satellite cloud extends to higher
halo masses, and a population of passive satellites appears at high stellar
masses. Most of the passive satellites are likely to be turned passive by
the stripping mechanisms acting in massive haloes. However, the highest passive
fractions are found at both high halo mass and stellar mass. These galaxies were
probably already passive when they were centrals and then merged with a more 
massive halo becoming passive satellites.
 
\subsection{Recovering predicted trends with observational proxies}

\begin{figure*}
\begin{center}
\includegraphics[width= 0.95\textwidth]{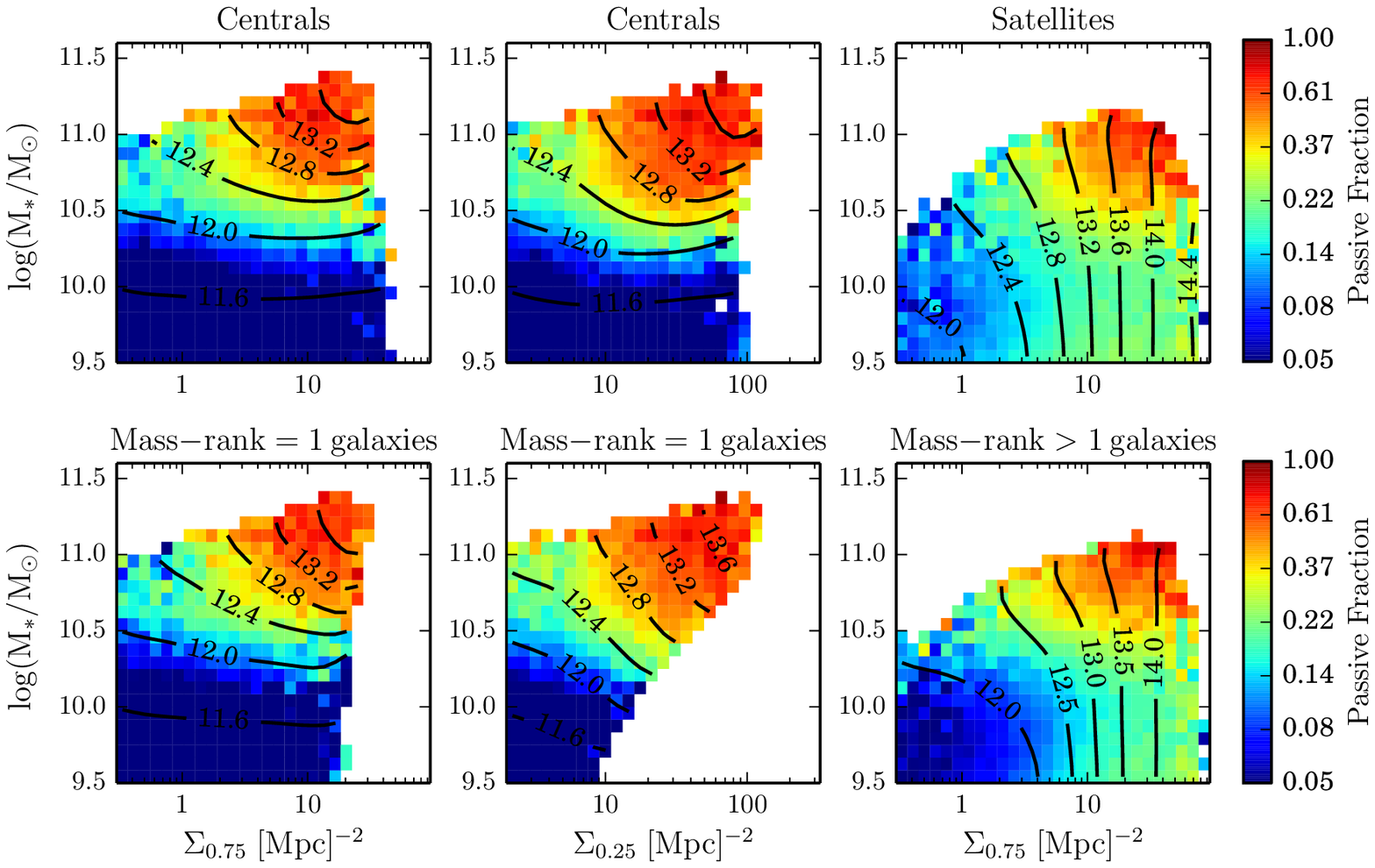}
\end{center}
\caption{Fraction of passive galaxies as a function of density on the scale $\Sigma_{0.75}$ and stellar mass for central galaxies (left panels), on the scale $\Sigma_{0.25}$ 
for centrals (central panels) and on the scale $\Sigma_{0.75}$ for satellite galaxies (right panels), at $z=1.08$.  The central/satellite definition is the one provided by the SAMs 
in the top panels and the one defined by the mass rank in the adaptive aperture in the bottom panels. The contours are drawn from the median halo mass in the same bins.} 
\label{PassfracenvMs}
\end{figure*}

In this section we investigate if, and how well, the predicted trends
of passive fraction as a function of halo mass and stellar mass can be
recovered using only observable quantities. We make use of the density
of galaxies in fixed apertures and the choice of centrals and
satellites is performed both using the model definition and the
observational mass rank method presented in Section \ref{RankCenSat}. 
We recall that the densities, number density contours and 
the median halo mass values presented in this section are obtained 
after application of the statistical weights described in Sections 
\ref{ModelSEL}, and \ref{methodENV}.

In Figure \ref{PassfracenvMs} we show the fraction of passive galaxies
(at $z=1.08$) as a function of density and stellar mass for centrals
(left and middle panels) and satellites (right panels). The top row
makes use of the separation between these two types as coded in the
models, while the bottom row uses the mass rank in the adaptive
aperture in order to divide the two types. The contours describe the
median halo mass in the same bins of stellar mass and density. Again,
each bin has to contain at least 10 objects for the passive fraction to
be computed.  A close examination of the direction of change
for passive fraction relative to both the axes and to the direction
of change for median halo mass, allows us to evaluate how well we
can use the parameters to track the trends seen in the pure model 
space (see Figure~\ref{PassfracMhMs}).

Let us start with the centrals. We know from Figure \ref{PassfracMhMs} 
that the passive fraction increases primarily with halo mass, and that at fixed 
halo mass there is if anything a slight anti-correlation with stellar mass. 
Can we see this in the observational parameter space? 

In the left panels in Figure \ref{PassfracenvMs} the density has been
computed on scales of 0.75 Mpc, which covers a super-halo scale
for all the haloes of mass below $10^{14} M_\odot$. 
To examine mostly intra-halo scales, we also examine the density
computed on the smallest scale (0.25 Mpc, middle panels).  

From the top panels, and without the halo mass contours, it would appear that
the stellar mass is the main driver of the correlation with passive fraction. 
Indeed this confusion is caused by the wide range of density seen
at low stellar mass: densities are reached which are just as large as
those for the centrals of high mass haloes (even on 0.25 Mpc scales).
as discussed in Section~\ref{Sect_Halomassenv}. 
However, when the halo mass contours are compared to the direction of increase of
the passive fraction, it is evident that the halo mass is the
main driver of the trend.

In the models, the low mass galaxies at high density show a passive fraction 
that is very similar to those at low density and comparable mass.
However in the real Universe those galaxies may
have already experienced physical processes driven by massive haloes, even
if they are currently outside the virial radius --
i.e. the ``backsplash'' population discussed by
\citet{mamon+04, balogh+00, ludlow+09, bahe+13, wetzel+13, hirschmann+14}.
Indeed, observational data suggests they behave more like satellites
\citep{wetzel+13, hirschmann+14}.
This implies that when the method is applied to an observational dataset,
it is useful to ``clean'' the sample of these high density, low mass objects. 
Fortunately, this happens as a direct consequence of making a mass rank 1 
selection as in the bottom panels. The adoption of the adaptive aperture 
means that our mass rank 1 galaxies are the most massive within an aperture not larger
than 0.75 Mpc but not smaller than 0.28 Mpc for galaxies in our
stellar mass range. Galaxies at low mass and high 0.25 Mpc density
are inevitably {\emph not} the most massive within this aperture, and
are excluded.  However this correlation depends on both density and 
stellar mass because the halo mass for central galaxies depends on both these parameters.


\begin{figure*}
\begin{center}
\includegraphics[width= 0.95\textwidth]{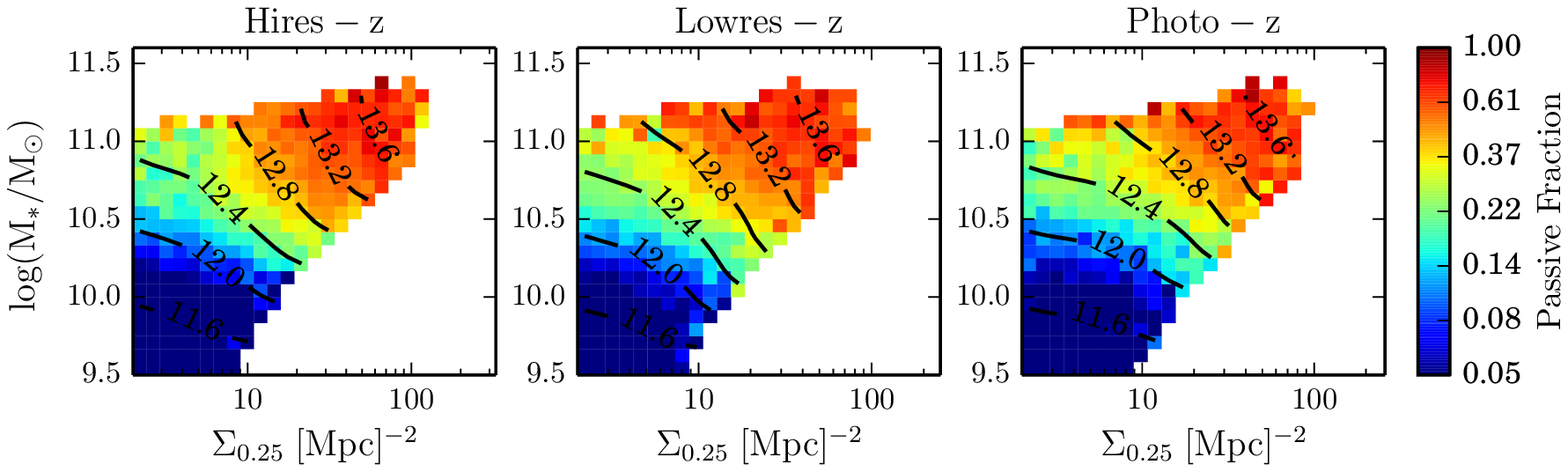}
\end{center}
\caption{Fraction of passive galaxies as a function of density on the scale $\Sigma_{0.25}$ and stellar mass 
for mass rank 1 galaxies (in the adaptive aperture) using hires-z (left panel), lowres-z (central panel), 
and photo-z (right panel) at $z=1.08$. The contours are drawn from the median halo mass in the same bins.} 
\label{PassfracenvMsaccuracy}
\end{figure*}

For satellites (which dominate in mass and number within the rich
haloes), the correlation between halo mass and density is very good
irrespective of the aperture used (the contours in the right
panels of Figure \ref{PassfracenvMs} are essentially vertical). 
Therefore the passive fraction trends as a function of halo
mass and stellar mass are easy to qualitatively recover
using the density and either the satellite definition
in the SAMs or the one coming from the mass rank. The low mass - high density
central population removed by the mass rank method ends up in the cloud of 
satellites. While those galaxies dominate the low mass centrals at high density,
they do not contribute much to the satellites at the same stellar mass and density.  
Density computed on either 0.25 or 0.75 Mpc work well in this regard. 

A general conclusion is that a well calibrated halo mass dependence on the 
observed properties (stellar mass, density) is crucial in understanding 
which physical properties are shaping the trends (in this case the passive 
fraction) we observe. 
While the results can be already be achieved with the SAM definition of 
centrals and satellites, it is important to stress that the use of the 
observationally motivated mass rank method provides the same result.

Finally, Figure \ref{PassfracenvMsaccuracy} shows how the passive fraction trends 
change if less accurate redshifts are used, as in our lowres-z and photo-z samples. 
We restrict ourselves to central galaxies defined
with the mass rank method in the adaptive aperture and density computed on the 0.25 Mpc scale.
Decreasing the quality of the redshift survey, the density - stellar mass 
correlation is less tight and the number of galaxies at the highest densities is reduced. 
The conclusions that can be drawn in this parameter space are unchanged. However, we 
stress that this conclusion comes with a number of caveats. First, the density dynamic range
is reduced when photo-z are used on all scales, but less so for 0.25 Mpc. This small scale can only
be used if the galaxy sampling is good enough, e.g. the stellar mass limit is low as in this 
work. Second, the use of the mass rank method ``cleans'' the sample
of low mass centrals which are projected in high density regions due to the less accurate 
photo-z, allowing us to obtain a trend similar to that for hires-z. Moreover, the density 
field is well reconstructed only with very good photometric redshifts. The photo-z accuracy 
typically depends on the number of photometric data points and their distribution 
across the rest-frame galaxy spectral energy distribution, generally being worse for 
fainter objects. Therefore, we warn the reader that the performance of photo-z in recovering 
the true density field should be carefully assessed for each individual sample, along with
the significance of any particular result based on this approach.

\section{Conclusions} \label{conclusions} 

In this work we have
characterised the definition of ``galaxy environment'' by means of the
projected density within fixed apertures at $z\sim$ 1-2. We
have tested our methods by applying them to the semi analytic models of
galaxy formation presented by \citet{guo+13} and based on a new run
of the Millennium simulation. We have focused on the correlation
between observables (density, stellar mass rank) and properties
provided only by the models (halo mass, central/satellite status). Then
we have studied to what extent our tools can recover the environmental
trends imprinted in the models in the context of the quenching of
centrals and satellites, extending to higher redshift the results of
\citet{hirschmann+14}. Our results can be summarised as follows:
    
\begin{enumerate}
\item The correlation between density and halo mass is not trivial and
a variety of effects are on stage at the same time. We find that
density poorly correlates with halo mass for centrals. This effect is
caused by the well defined boundaries of haloes in the SAMs at high density:
galaxies within those boundaries are satellites hosted by a high-mass
halo, while those outside are central galaxies of low-mass haloes. 
It has been shown by \citet{hirschmann+14} that
density correlates with halo mass only for massive centrals. For all
centrals at fixed density, the distribution of halo mass broadens 
so much that the density-halo mass correlation is lost. This is 
consistent with similar results by \citet{woo+13}. 
On the other hand density correlates well with halo
mass for satellites, irrespective of the aperture used.
\item Central galaxies in the accretion regions of massive haloes can
  be highlighted with a simple but effective method. We replaced the
  nominal halo mass with that of the most massive halo within a
  physical (3D) distance of 1 Mpc. This traces the dominant DM mass
  nearby and we recover a correlation between density and this
  halo-mass for centrals which is similar to that for the satellites.
\item The stellar mass rank is an effective method to identify centrals 
and satellites. We have parameterized the performance of this method 
in terms of purity and completeness of the mass rank identification with 
respect to the SAM definition. We have tested various apertures where the 
rank is computed. For central galaxies we find that the larger the aperture, 
the higher is the purity but the lower is the completeness. 
In order to improve both the completeness at low halo masses
and the purity in massive haloes we have defined an adaptive aperture that 
depends on the stellar mass of the galaxy. This 
method is as good as a fixed 0.75 Mpc aperture in terms of purity but with 
an improvement of $\sim 30\%$ in completeness at halo masses below $10^{13} M_\odot$.  
The method is not strongly sensitive to the stellar mass limit or the
spectroscopic sampling rate, though less so for the completeness of satellites. 
Our results for purity and completeness are remarkably consistent with 
\citet{knobel+12}, despite the different method and sample selection.
\item A strong $M_*$-$M_{\rm halo}$ correlation is predicted by the
models for central galaxies. Passive centrals dominate above $M_{\rm
halo} = 10^{12.5} M_\odot$ due to strong AGN feedback, correlated to bulge
growth \citep{wilman+13}.  However, the density-halo mass
correlation for central galaxies is far from being linear 
or independent of stellar mass. Therefore the recovered trends do not
only depend on density but also on stellar-mass. \emph{Within
a purely observational parameter space, we are able to recover
these trends.} This requires three steps:
\begin{itemize}
\item{A careful identification of central (mass rank 1) and
      satellite (mass rank $>1$) galaxies. To achieve high completeness
      of central galaxy identification, we have applied an adaptive
      aperture. For the interpretation of observations, we would
      ideally exclude centrals living close to massive haloes as
      backsplash galaxies can complicate the physical interpretation of
      central galaxies.}
\item{A calibration showing how the halo mass depends on density and stellar 
mass, for a population of mock galaxies selected in exactly the same way as in 
observations.}
\item{Ideally (if the sampling and depth are suitable), density
      should be computed on scales comparable to the aperture used to
      identify central galaxies. This ensures a cleaner correlation
      between density and halo mass for central galaxies.} 
\end{itemize}

The redshift accuracy does not negatively impact on this result. 
However, such a conclusion requires a combination of good photometric redshifts, 
deep survey limits, and the mass rank method to identify centrals and satellites. 
\end{enumerate}

Finally we describe a possible way of using these results to 
understand the environmental trends in observational data. First of 
all the sample selection in the models should be as close as possible
to that in the data. The model galaxies need to be weighted to match 
the mass (and possibly also the magnitude and colour) distributions.
Then the quantification of densities needs to take into account the 
redshift accuracy of the survey under investigation. At this point the
density distributions of real and model data can be compared. The models
then provide calibrations of properties such as halo mass which can be 
contrasted with observed properties such as
passive fraction (see Figure \ref{PassfracenvMs}). This will help 
identifying the physical processes driving the trends.

\section*{acknowledgements}
MF and DJW acknowledge the support of the Deutsche Forschungsgemeinschaft 
via Project ID 387/1-1.  FF acknowledges financial contribution from the
grants PRIN MIUR 2009 ``The Intergalactic Medium as a probe of the growth of 
cosmic structures'' and PRIN INAF 2010 ``From the dawn of galaxy formation''.
GDL, MH and EC acknowledge financial support from the European Research Council
under the European Community's Seventh Framework Programme (FP7/2007-2013)/ERC
grant under agreement n. 202781. PM has been supported by a FRA2012 grant of 
the University of Trieste and PRIN MIUR 2010-2011 J91J12000450001 ``The dark 
Universe and the cosmic evolution of baryons: from current surveys to Euclid''. 

MF thanks Gerard Lemson for help and advices in 
handling the G13 models, and Olga Cucciati, Michael Balogh, and Michael Cooper for
comments which helped to improve the manuscript. We thank the anonymous referee
for his/her comments that helped improving the quality of the paper.

\appendix

\section{A multi-scale approach}\label{sec:multiscales}
\begin{figure*}
\begin{center}
\includegraphics[width= 0.95\textwidth]{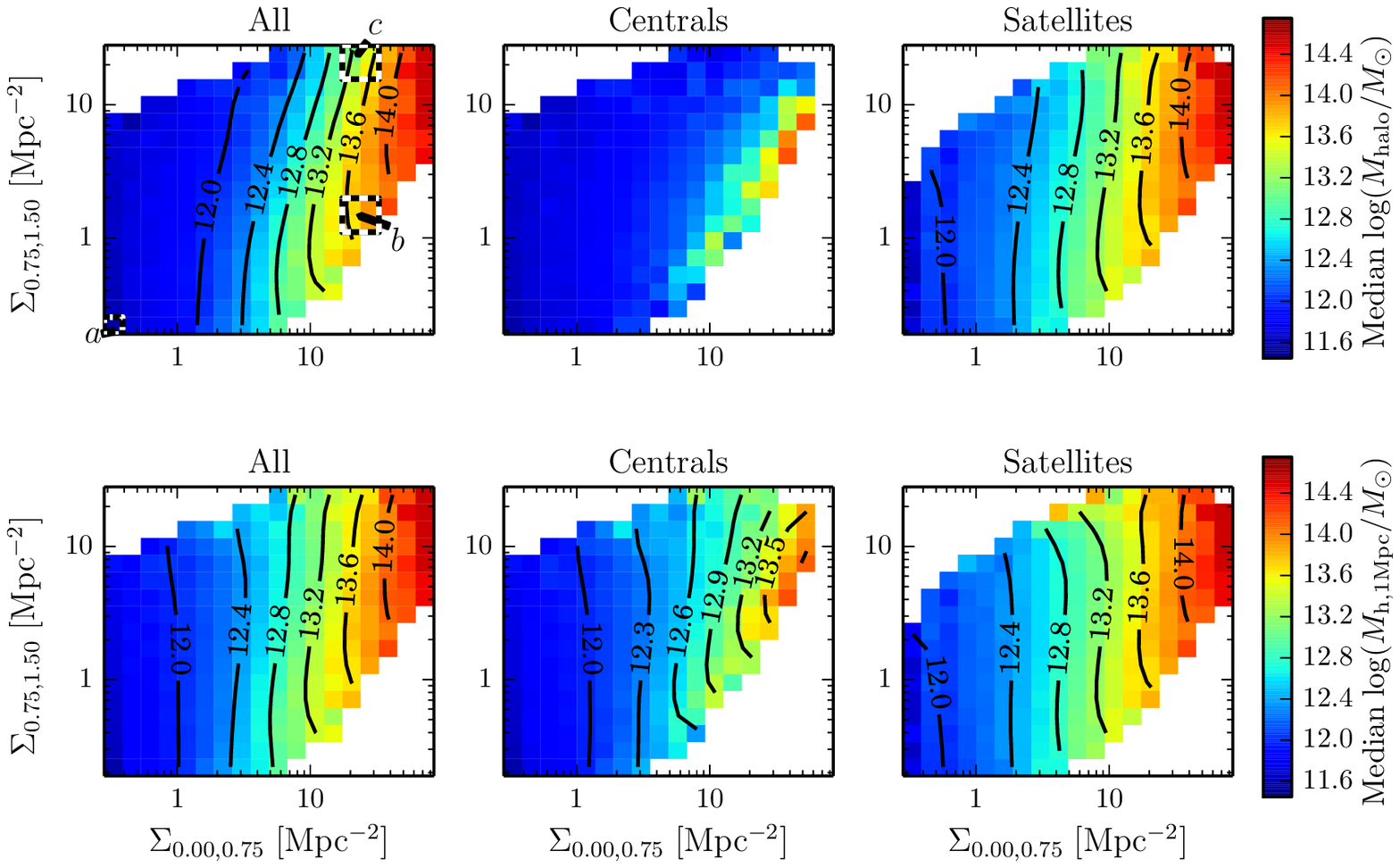}
\end{center}
\caption{Top panels: halo mass dependence with density on scales $\Sigma_{0.00,0.75}$ and $\Sigma_{0.75,1.50}$ for all galaxies (left panels), central 
galaxies only (central panels) and satellite galaxies only (right panels) at $z= 1.08$. Bottom panels: as before but the nominal halo mass has been replaced with that of
the most massive halo within a 3D physical sphere of 1Mpc radius ($M_{\rm{h,1Mpc}}$). The labels $a$, $b$, and $c$ in the top left panel highlight three bins 
(black and white squares) whose halo mass distributions are shown in Figure \ref{Halobin2Ddens}.  } 
\label{Halomassenv}
\end{figure*}

As we already discussed in Sect. \ref{Sect_Halomassenv}, the super-halo scale imprints a complex
dependence on the halo mass vs density correlation for central
galaxies. Here we show how the combination of two scales of density 
can be used to identify low mass galaxies at high density. 
In Figure \ref{Halomassenv} (top panels) the median halo mass dependence 
on density is shown for a pair of independent scales: $\Sigma_{0.00,0.75}$ 
and $\Sigma_{0.75,1.50}$. The smaller scale is chosen to cover $\sim 2 - 3$
times the virial radius of haloes more massive than $10^{13} M_\odot$,
while the larger aperture highlights the second order effects on
super-halo scales.  The left hand panels include all galaxies while
those in the middle and on the right include only centrals and
satellites respectively. Over-plotted contours, where they exist, are 
computed from a smoothed map of the data using a Gaussian filter with 
$\sigma = 1.5$ pixels.
Smoothing is required to wash out the local features while keeping the
overall direction of change of the halo mass with density.  Indeed,
contours in the top left panel are typically aligned with the vertical
axis. Indeed halo mass correlates more strongly with smaller
scale densities than with the larger ones.

A closer look shows that the contours are not fully aligned with the
large scale-axis. There is a clear anti-correlation with large-scale
density at fixed small-scale density, i.e. the median halo mass
decreases when increasing the large scale density at \emph{fixed} small
scale density.  The top middle panel shows that the median halo
mass is not dependent on density for central galaxies except for the
extremely high small-scale densities.  Finally, the top right panel
shows that the median halo mass for satellite galaxies depends
almost entirely on the small-scale density, as seen in Figure
\ref{Halomass1Denv}.

\begin{figure*}
\begin{center}
\begin{tabular}{c c c}
\hspace{2cm} \includegraphics[scale = 0.28]{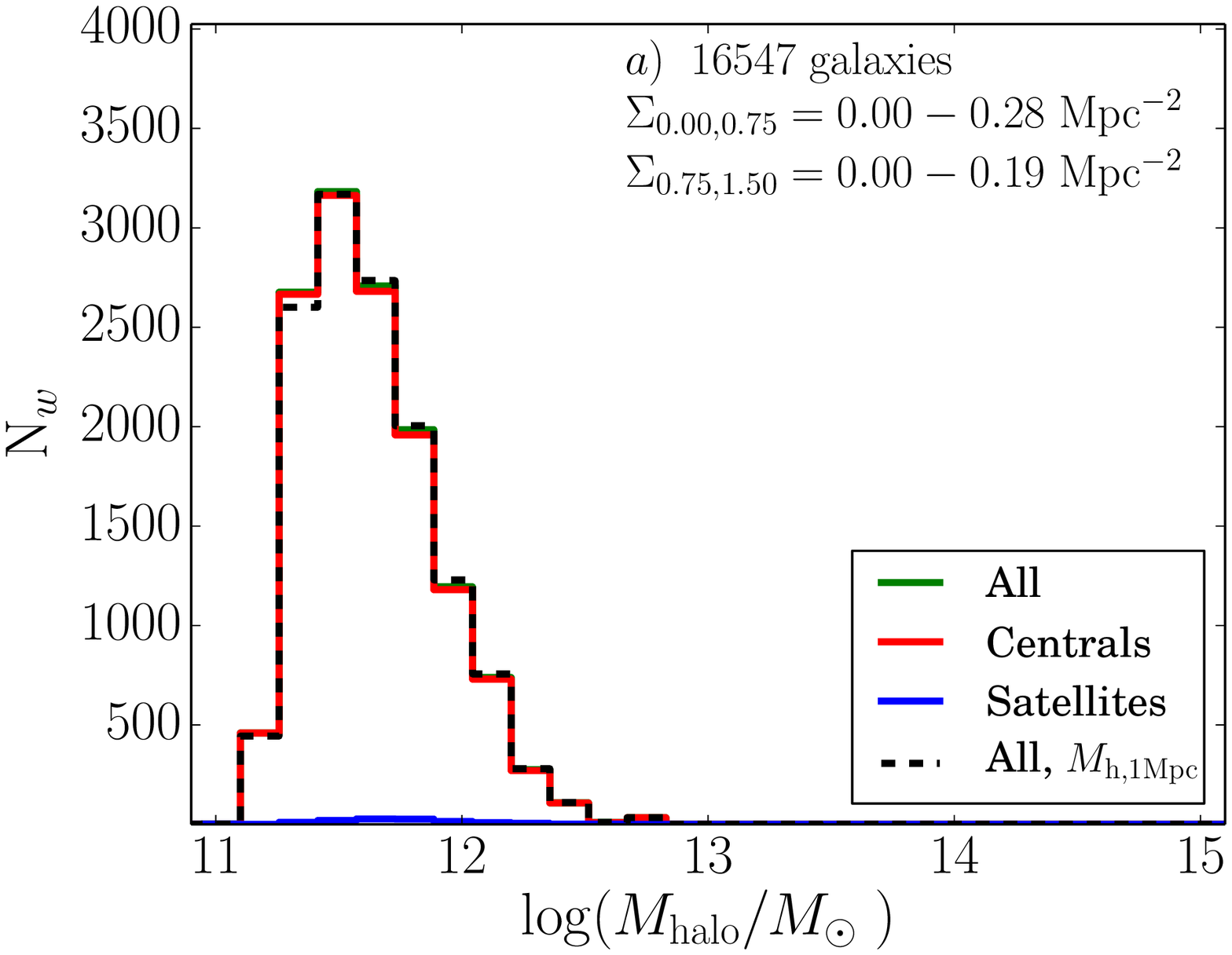}  & \includegraphics[scale = 0.28]{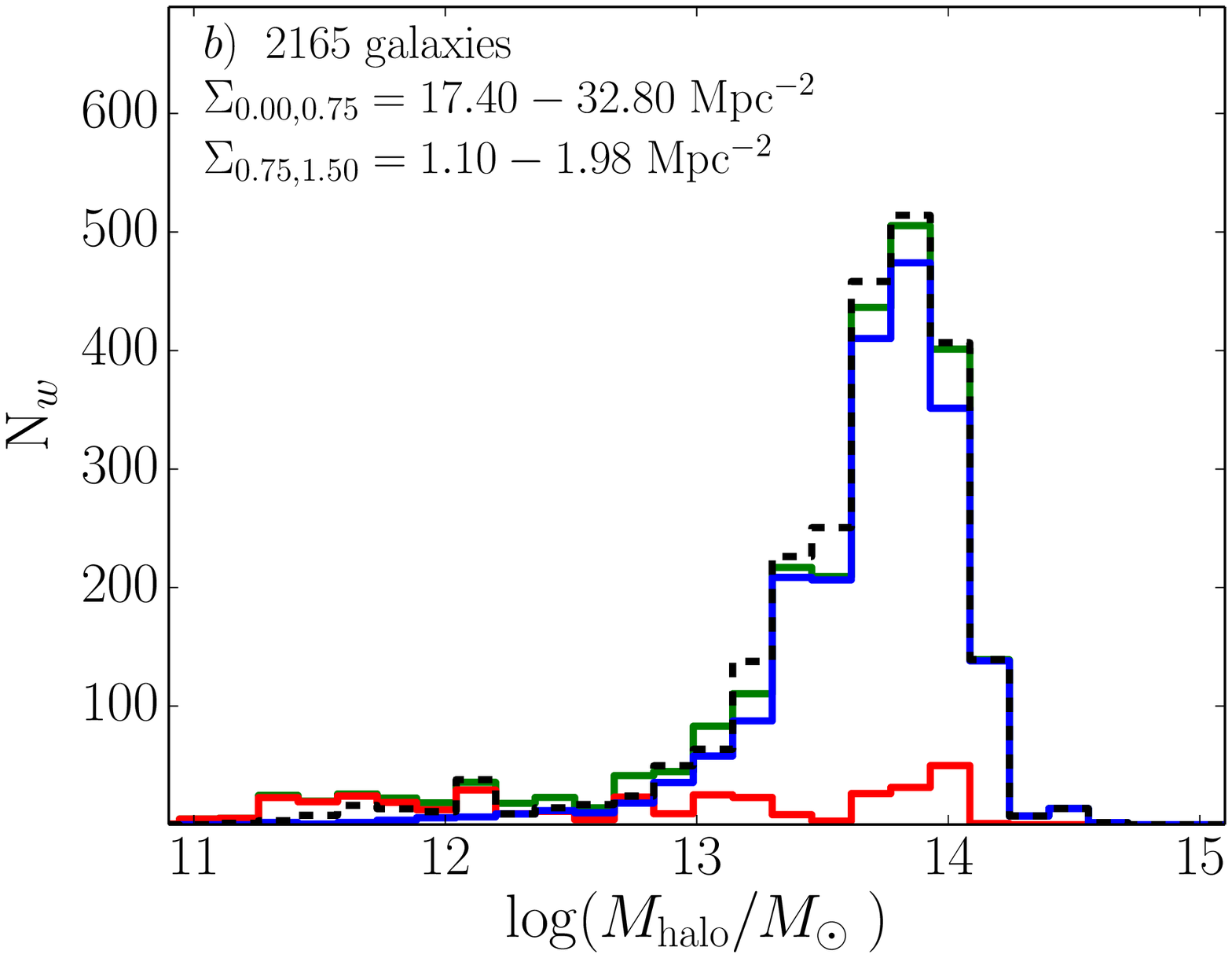}  & \includegraphics[scale = 0.28]{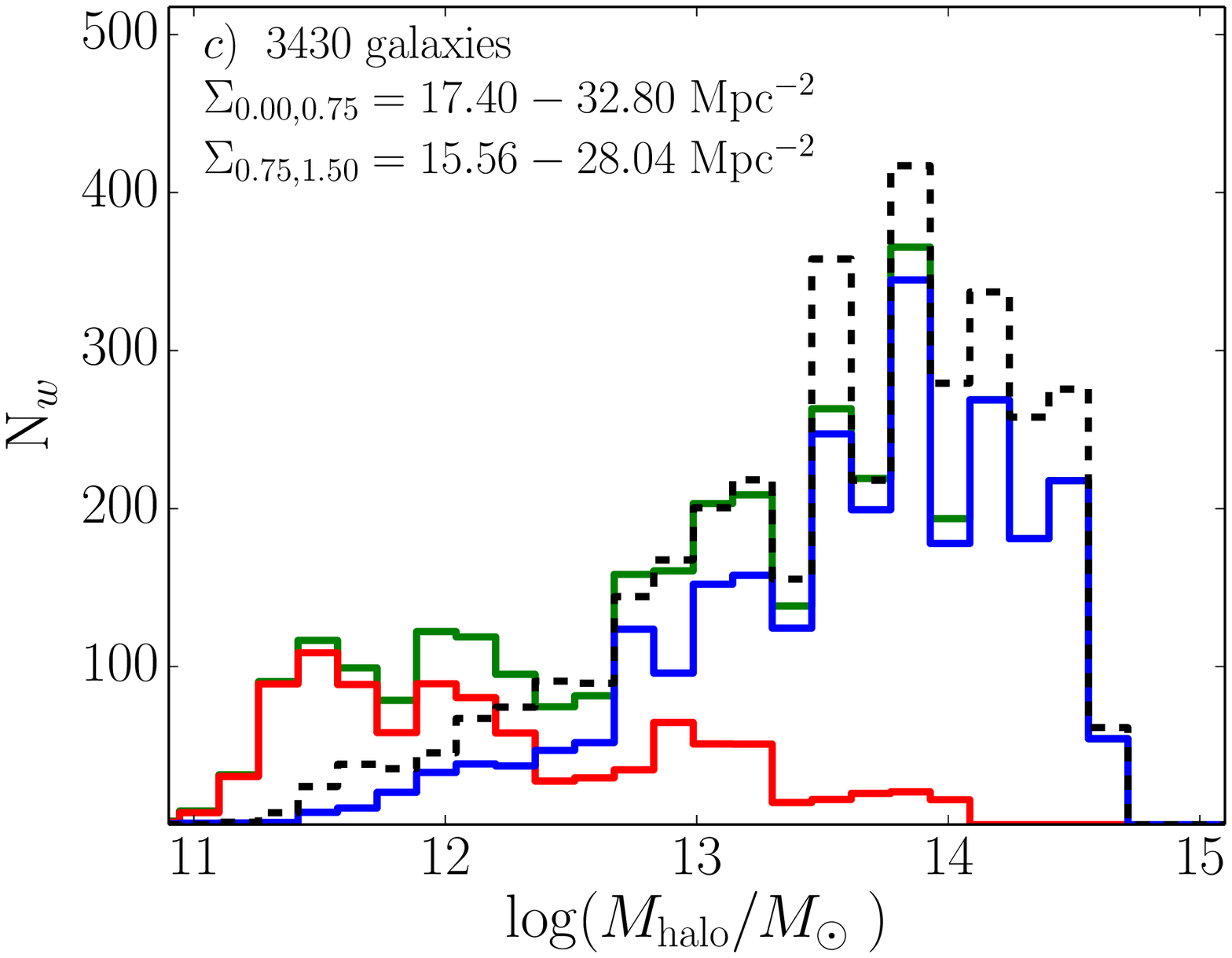} \\
\end{tabular}
\end{center}
\caption{Halo mass distribution in three bins indicated using black and white squares and labelled $a$, $b$, and $c$ in 
the top left panel of Figure \ref{Halomassenv}. } 
\label{Halobin2Ddens}
\end{figure*}

In order to understand these patterns it is important to roughly sketch
the densities experienced by galaxies living in the core or in the
outskirts of their own halo. A galaxy living in the centre of its own
halo has high densities within annuli up to the size of the halo and
low densities beyond.  A galaxy living just beyond the halo virial
radius has an intermediate density on small scales (as the aperture
encompasses a fraction of the halo) and a high density on the larger 
scale, since the nearby halo core is located
in this annulus. If we consider that those galaxies beyond the halo
boundary are considered centrals, we fully understand why the density
does not correlate positively with halo mass on large scales.

For the bottom panels of Figure \ref{Halomassenv} we show instead the 
mass of the most massive halo within 1 Mpc of each galaxy, $M_{\rm{h,1Mpc}}$. 
In contrast to the median mass of the parent halo, the mass of the most 
massive nearby halo correlates strongly with density for centrals 
 (middle bottom panel). As already stressed in the text, the satellite galaxies 
are almost unaffected. The bottom left panel shows how the complete population
behaves.  The contours are now essentially vertical and the
anti-correlation of halo mass with large-scale density at fixed
small-scale density disappears.

We select three representative regions ($a$, $b$, $c$) in the upper
left panel in Figure \ref{Halomassenv} and we study the distributions of 
halo mass in these regions in Figure \ref{Halobin2Ddens}).

The left-hand panel ($a$) shows the distribution of halo masses in the
lowest density bin on both scales.  Almost all the galaxies are
centrals (red solid) living in low mass haloes and the halo mass
replacement has little effect on the overall distribution (blue solid)
as the majority of them have no neighbours within 1 Mpc. With the
Millennium-II simulation we get the same result and so this is robust
against resolution effects. The centre and right-hand panels ($b$ and
$c$) are chosen to have the same high density on the inner 0.75 Mpc
scale but very different densities in the outer annulus, highlighting
the importance of the large-scale density.  The galaxies whose
large-scale density is low (panel b) live near the centre of
their host halo, thus the sample is made almost entirely of satellites
(blue solid) and the effect of centrals on the overall distribution is
negligible. On the other hand, when $\Sigma_{0.75,1.50}$ is high
(panel c), the contribution from centrals having halo masses
smaller than $10^{12.5} M_\odot$ is 24\%, causing a decrease in
the median halo mass. This population of low halo mass centrals at
high density can contain a significant population of ``backsplash''
galaxies, as discussed in Section~\ref{Sect_Halomassenv}. We see that
the use of multiple scales can help identify such populations.
Figure~\ref{Halomassenv} also shows the distribution of
$M_{\rm{h,1Mpc}}$ (black dashed lines) for our three bins: in panel
c, the fraction of centrals with halo masses smaller than $10^{12.5}
M_\odot$ is reduced  to 10\%.  In this case the halo mass
distribution is skewed to higher halo mass becoming similar to that in
panel $b$.

\label{lastpage}

\end{document}